\documentclass[11pt,a4paper]{article}
\usepackage{jcappub}
\usepackage{graphicx}
\usepackage{float}
\usepackage{multirow}
\usepackage{booktabs}
\usepackage{threeparttable}
\usepackage{color}

\abstract{

We constrain two non-flat time-evolving dark energy cosmological models
by using Hubble parameter data, Type Ia supernova apparent magnitude
measurements, and baryonic acoustic oscillation peak length scale observations. The
inclusion of space curvature as a free parameter in the analysis results 
in a significant broadening of the allowed range of values of the parameter
that governs the time evolution of the dark energy density in these models.
While consistent with the ``standard" spatially-flat $\Lambda$CDM cosmological model,
these data are also consistent with a range of mildly non-flat, slowly time-varying dark 
energy models. After marginalizing over all other parameters, these data require the
averaged magnitude of the curvature density parameter $|\Omega_{k0}| \lesssim 0.15$ 
at 1$\sigma$ confidence.
}

\begin{document}
\title{Observational constraints on non-flat dynamical dark energy cosmological models}

%
%

%

\author[a,1]{Omer Farooq,\note{Corresponding author.}}
\author[a,b]{Data Mania,}
\author[a]{Bharat Ratra}


\affiliation[a]{Department of Physics, Kansas State University,  116 Cardwell Hall, Manhattan 66506 USA}
\affiliation[b]{Center for Elementary Particle Physics, Ilia State University, 3-5 Cholokashvili Ave., Tbilisi 0179, Georgia}

\emailAdd{omer@phys.ksu.edu}
\emailAdd{mania@phys.ksu.edu}
\emailAdd{ratra@phys.ksu.edu}

\date{ \today}

\maketitle
\section{Introduction}

There is significant observational evidence that the Universe
is currently undergoing accelerated expansion.
Most cosmologists believe that dark energy dominates the current cosmological
energy budget and is responsible for this accelerated 
expansion  \citep[for reviews of dark energy see][and references therein]
{li2012,tsujikawa2013,sola2013}.\footnote[4]{
Some instead argue that these observations should be viewed as an indication 
that general relativity needs to be modified on these large cosmological
length scales. For recent reviews of modified gravity see \cite{Capozziello2011}, 
\cite{trodden2012}, and
references therein. We assume here that general relativity provide an accurate description
of gravitation on cosmological length scales.} In addition, if one assumes that the dark energy density is close to or time independent, 
cosmic microwave background (CMB) anisotropy measurements indicate that the Universe must be close to 
or spatially flat (\citealt[][]{ade13} and references therein; for an early indication see \citealt[][]{Podariu2001b}).
Conversely, if one assumes that
the space sections are flat, the data favor a time-independent cosmological constant. However,
as far as we are aware, there has not been an analysis of observational data based 
on a physically consistent non-spatially-flat dynamical dark energy model. In this paper we present the first
such analysis.

As a warm-up exercise, we consider the popular XCDM parameterization of dynamical dark energy.\footnote[5]{Here
dark energy is taken to be a spatially-homogeneous $X$-fluid with a time-evolving energy density that 
dominates the current cosmological energy budget, with non-relativistic cold dark matter (CDM) being the next
largest contributor.} 
We generalize this parameterization
to the case when the spatial hypersurfaces have non-zero curvature, a case that has been 
considered previously, along with more general fluid parameterizations \citep[see e.g.,][]{crooks2003,
ichikawa2006,ichikawajcap2006,zhao2007,wang2007,ichikawajcap2007, 
gong2008,ichikawajcap2008,virey08,mortonson2009,dossett2012}. These authors \citep[also see][]{write2006}
emphasize the important point that in this case there is a degeneracy between the spatial curvature and the 
$X$-fluid equation of state parameter, and so cosmological data are not as constraining in this case compared to the 
case when either spatial curvature vanishes or the dark energy density is a constant. However, the XCDM parameterization 
is incomplete, as it does not describe spatial inhomogeneities \citep[see e.g.,][]{ratra91, podariu2000}.

The XCDM parameterization is a generalization of the standard $\Lambda$CDM cosmological 
model \citep[][]{peebles84}. In the $\Lambda$CDM case the current energy budget is dominated by a time-independent 
cosmological constant $\Lambda$. It is well-known that the $\Lambda$CDM model has some puzzling 
features which are more easily understood if, instead of remaining constant like $\Lambda$, the 
dark energy density gradually decreases with time.\footnote[6]{Note
that there also are tentative observational indications that the standard CDM structure formation model, which is assumed in
the $\Lambda$CDM cosmological model, might need to be improved upon \citep[][and references therein]{Peebles&Ratra2003,Weinberg2013}.}

The simplest, complete and consistent time-varying dark energy 
model is $\phi$CDM \citep{Peebles&Ratra1988, 
Ratra&Peebles1988}.\footnote[7]{For recent discussions of other time-varying 
dark energy models see \cite{liu2012}, \cite{Garcia-Salcedo2012}, \cite{Benaoum2012}, 
\cite{Ayaita2012}, \cite{Ferreira2013}, \cite{Bezrukov2013}, \cite{Liao2013}, and references therein.}
Here dark energy is modeled as a scalar 
field, $\phi$, with a gradually decreasing (in $\phi$) potential energy 
density $V(\phi)$. In this paper we assume an inverse-power-law potential energy 
density, $V(\phi) \propto \phi^{-\alpha}$, where $\alpha$ is a nonnegative 
constant \citep{Peebles&Ratra1988}. At $\alpha = 0$ the
$\phi$CDM model reduces to the corresponding $\Lambda$CDM case. 
The $\phi$CDM model was originally formulated in a spatially-flat cosmological model. In this paper we consider 
the \cite{Anatoly2} generalization of the $\phi$CDM model to non-flat space.\footnote[8]{Curved-space scalar field dark energy models have been studied in the past \citep[see e.g.,][]{Aurich2002,Aurich2003,Aurich2004,Thepsuriya2009,Chen&Guo2012,Gumjudpai2012}. However, as far as we are aware, \cite{Anatoly2} were the first to establish that the scalar field solution is a time-dependent fixed point or attractor even in the curvature dominated epoch.}

For some time now, most observational constraints have 
been reasonably consistent with the predictions of the ``standard'' 
spatially-flat 
$\Lambda$CDM model \citep[for early indications see e.g.,][]{jassal10,
wilson06,Davis2007,allen08}. The big four, CMB anisotropy \citep[e.g.,][]{ade13}, supernova Type Ia (SNIa)
apparent magnitude versas redshift \citep[e.g.,][]{suzuki12,Salzano2012,
Campbell2013}, baryonic acoustic oscillation (BAO) peak
length scale \citep[e.g.,][]{Percival2010,beutler2011,blake11,Mehta2012}, and Hubble parameter 
as a function of redshift \citep[e.g.,][]{Chen2011b,moresco12,Busca12,farooq3} measurements
provide the strongest support for this conclusion. 

Other measurements that have been used to constrain cosmological 
parameters include, for example, galaxy cluster gas mass fraction as a function of redshift 
\citep[e.g.,][]{allen08, Samushia&Ratra2008, tong11, Lu2011,Landry2012}, 
galaxy cluster and other large-scale structure properties 
\citep[][and references therein]{mortonson2011, 
Devi2011, Wang2012, deboni13,Batista2013}, gamma-ray burst luminosity distance as a function of 
redshift \citep[e.g.,][]{Samushia&Ratra2010, Wang2011, Busti2012,Pan2013}, 
HII starburst galaxy 
apparent magnitude as a function of redshift \citep[e.g.,][]{plionisetal10, 
plionisetal11, Mania2012}, angular size as a function of redshift 
\citep[e.g.,][]{guerra00, Bonamente2006, Chen2012}, and strong 
gravitational lensing \citep[][and references therein]{chae04, lee07, 
biesiada10, Suyu2013}.\footnote[9]{
Future space-based SNIa and BAO-like measurements 
\citep[e.g.,][]{podariu01a, Samushia2011, Sartoris2012, Basse2012, Pavlov2012},
as well as measurements based on new techniques 
\citep[][and references therein]{Appleby2013, Arabsalmani2013}
 should soon provide interesting constraints on cosmological
parameters.}
 While the constraints from 
these data are typically less restrictive than
those derived from the $H(z)$, SNIa, CMB anisotropy, and BAO data, both types of
measurements result in largely compatible constraints that generally 
support a currently accelerating cosmological expansion. This provides
confidence that the broad outlines of a ``standard'' cosmological
model are now in place. 

In this paper we consider an extension of this ``standard" cosmological 
model by allowing for the possibility of non-zero space curvature. As mentioned
above, we consider two possibilities, a generalization of the XCDM parameterization
as well as the \cite{Anatoly2} generalization of the $\phi$CDM model. In this 
paper we derive constraints on the parameters of these options by using $H(z)$,
SNIa, and BAO data. This work extends previous 
work \cite[e.g.][]{Farooq2012,Farooq2013a,farooq3} via the inclusion of space curvature
in dynamical dark energy models.

Here we do not make use of the last of the big four data, that of CMB anisotropy.
While CMB anisotropy data are widely credited with providing the strongest evidence 
for a very small contribution to the current energy budget from spatial
curvature (however, see discussion above), it is not straightforward to include them 
because they require an analysis of the 
evolution of spatial inhomogeneities. In the case of the XCDM
parametrization this is not possible without an ad hoc extension. In the $\phi$CDM
case this requires a detailed computation, including the assumption of an early epoch of inflation
in non-flat space and a derivation of the concomitant power spectrum needed for the CMB anisotropy
computation. It is well known that the observed CMB anisotropy is almost certainly a consequence of 
quantum-mechanical zero-point fluctuations generated during inflation \citep[see e.g.,][]{Fischler1985,ratra92}.
While conceptually similar, the computation of the primordial spectrum is 
more involved in the spatially-curved case than in the flat model \cite[see e.g.,][]{Gott1982,Ratra&Peebles1994,Ratra&Peebles1995,
Kamionkowski1994,Bucher1995,Lyth1995,Yamamoto1995,Ganga1997,gorshi1998}. Consequently, the computation
of CMB anisotropy constraints is beyond the scope of this initial paper. Even though we do not
use CMB anisotropy constraints here, a combination of the other three of the big four data--- $H(z)$,
SNIa, and BAO--- results in reasonably tight constraints on space curvature. For technical
computational reasons we believe our XCDM parametrization constraints are more reflective of the true
constraints on space curvature.\footnote[10]{It is much more time consuming
 to do the $\phi$CDM computation, so we assumed a
narrower prior on space curvature in this case, which we suspect leads to slightly tighter but less reliable
constraints.} In the XCDM case, marginalizing over all other parameters, the $H(z)$, SNIa, and BAO data
require $|\Omega_{k0}|\leq 0.15$ and 0.3 at about 1$\sigma$ and 2$\sigma$ confidence.

Our paper is organized as follows. In Sec.\ {\ref{sec:Models}} we
present the basic equations of the $\phi$CDM dark energy model and the XCDM dark energy parameterization. Constraints from observational data are derived in Sec.\
{\ref{sec:Constraints}}. We conclude in 
Sec.\ {\ref{summary}}.

\label{intro}

\section{Time-varying dark energy models in curved space}
\label{sec:Models}

In this section we summarize the two models we constrain. These are the \cite{Anatoly2}
generalization to curved space of the time-evolving dark energy $\phi$CDM model 
\citep{Peebles&Ratra1988, Ratra&Peebles1988}, as well as the curved space
generalization of the widely-used XCDM
dynamic dark energy parameterization in which dark energy is 
modeled as a spatially-homogeneous time-dependent $X$-fluid. 

We assume that general relativity provides an accurate description
of gravitation on cosmological scales. The equations of motion are 
Einstein's field equations,
\begin{equation}
R_{\mu\nu} - \frac{1}{2}Rg_{\mu\nu} = 8 \pi G T_{\mu\nu} - \Lambda g_{\mu\nu}.
\end{equation} 
Here $R_{\mu\nu}$ and $R$ are the Ricci tensor and scalar, 
$g_{\mu\nu}$ is the metric tensor, $\Lambda$ is the cosmological constant,
$T_{\mu\nu}$ is the energy-momentum tensor of the matter present,  and 
$G$ is the Newtonian gravitational constant.

At late times we can ignore radiation and model non-relativistic (cold dark and
baryonic) matter as a perfect fluid with 
energy-momentum tensor $T_{\mu\nu} = 
{\rm diag} (\rho, p, p, p)$ where $\rho$ and $p$ are the energy density 
and the pressure of the fluid. Assuming the cosmological principle of
spatial homogeneity, Einstein's equations reduce to 
the two independent Friedmann equations, 
\begin{equation}
\label{eq:F1}
\left(\frac{\dot{a}}{a}\right)^2 = \frac{8\pi G}{3}\rho-\frac{K^2}{a^2}+\frac{\Lambda}{3},
\end{equation}
\begin{equation}
\label{eq:F2}
\hspace{1 cm} \frac{\ddot{a}}{a}\ \ =-\frac{4\pi G}{3} (\rho +3p)+\frac{\Lambda}{3}.
\end{equation}
Here $a(t)$ is the cosmological scale factor, which is the ratio of the physical 
distance to the co-moving distance of a sufficiently distant object (so that 
the spatial homogeneity assumption is valid), an over-dot denotes a 
derivative with respect to cosmological time, $K^2$ represents the curvature of 
spatial hypersurfaces (and can have three discrete values $-1$, $0$, or $+1$,
corresponding to hyperbolic, flat, and spherical geometry respectively), and $\rho$ and $p$
are the sums of all (time-dependent) densities and pressures of the various forms of matter
present. 

With a single type of matter, the Friedmann equations\ (\ref{eq:F1})---(\ref{eq:F2}) are two equations with 
three time-dependent unknowns, $a(t)$, $\rho(t)$, and $p(t)$. We can complete the system 
of equations with an equation of state for each type of matter. This
is a relation between pressure and energy
density for each type of matter,
\begin{equation}
\label{eq:EoS}
   p= p(\rho) = \omega \rho, 
\end{equation}
where $\omega$ is the dimensionless equation-of-state parameter for an ideal fluid.
For non-relativistic matter $\omega=0$, while
$\omega=-1$ corresponds to a standard cosmological constant $\Lambda$, and $\omega 
< -1/3$ corresponds to the XCDM parameterization.
 
Equations\ (\ref{eq:F1})---(\ref{eq:EoS}) form a closed set and can be used to derive 
the energy conservation equation, 
\begin{equation}
\label{eq:EC}
   \frac{\dot{\rho}}{\rho} =-3\ \frac{\dot{a}}{a}\ (1+\omega).
\end{equation} 
This first-order linear differential equation can be solved 
with the boundary condition $\rho(t_0)=\rho_0$, where $t_0$ is the current time
and $\rho_0$ is the current value of the energy density of the particular type of
matter under consideration. The solution is
\begin{equation}
\label{eq:EC1}
   \rho(t)=\rho_0 \left(\frac{a_0}{a}\right)^{3(1+\omega)},
\end{equation}
where $a_0$ is the current value of the scale factor. If there 
are a number of different species of non-interacting fluids, 
then Eq.\ (\ref{eq:EC1}) holds separately for each of them 
with the corresponding $\omega$ and $\rho_0$. For a non-relativistic 
gas (cold matter) $\omega=\omega_m=0$ and $\rho_m \propto a^{-3}$, 
for a homogeneous $X$-fluid $\omega=\omega_{X}<-1/3$ and $\rho_X\propto a^{-3(1+\omega_X)}$, 
and for spatial curvature $\omega=\omega_{K}=-1/3$ and $\rho_K\propto a^{-2}$. 

The ratio $\dot{a}(t)/a(t)$ in Eq.\ (\ref{eq:F1}) is the Hubble parameter
$H(t)$. The present value of the Hubble parameter is the Hubble constant 
$H_0$. To rewrite the Friedmann equation\ (\ref{eq:F1}) in
terms of observable parameters we define the dimensionless redshift $z = a_0/a - 1 $, and 
the present value of the density parameters,
\begin{equation}
\label{eq:density parameters}
   \Omega_{m0} = \frac{8\pi G\rho_{m0}}{3H_0^2}, \ \ \ \ \ \ \
   \Omega_{K0}=\frac{-K^2}{(H_0 a_0)^2}, \ \ \ \ \ \ \
   \Omega_{X0} = \frac{8\pi G\rho_{X0}}{3H_0^2}.
\end{equation}
Here we have parameterized dark energy as a spatially homogeneous $X$-fluid with 
current density parameter value $\Omega_{X0}$, $\Omega_{m0}$ is the current non-relativistic
(baryonic and cold dark) matter density parameter, and $\Omega_{K0}$ is that of
spatial curvature (with $\Omega_{K0}>0$ corresponding to an open or hyperbolic spatial geometry). 
With these definitions Eq.\ (\ref{eq:F1}) becomes
\begin{eqnarray}
\label{eq:XCDMwithc}
H^2(z; H_0, \textbf{p})\hspace{-2mm}&=&\hspace{-2mm}H_0^2 \left[\Omega_{m0} (1+z)^3 + (1-\Omega_{m0}-\Omega_{K0})(1+z)^{3(1+\omega_{X})}
                          + \Omega_{K0} (1+z)^2 \right], 
\end{eqnarray}
where we have made use of $\Omega_{X0} = 1-\Omega_{m0}-\Omega_{K0}$.
This is the Friedmann equation for the XCDM parameterization with non-zero 
spatial curvature. In this case the cosmological parameters are taken to be
$\textbf{p}=({\Omega_{m0},\omega_{X},\Omega_{K0}})$. The XCDM parameterization is incomplete, as it
cannot describe the evolution of energy density inhomogeneities \citep[see e.g.,][]{ratra91,podariu2000}.

The second model we consider is the simplest, complete and consistent dynamical dark energy model, 
$\phi$CDM, generalized to include non-zero spatial curvature \citep{Anatoly2}. 
In this case dark energy is modeled as a slowly-rolling scalar field 
$\phi$ with an, e.g., inverse-power-law potential energy density 
$V(\phi)=\kappa m_p^2 \phi^{-\alpha}/2 $ where $m_p=1/\sqrt{G}$ is 
the Planck mass and $\alpha$ is a non-negative parameter that 
determines the coefficient $\kappa$ \citep{Peebles&Ratra1988}. 
The scalar field part of the $\phi$CDM model action is
\begin{equation}
   S=\frac{m_p^2}{32\pi}\int{\sqrt{-g}\left( ~g^{\mu \nu}
   \partial_\mu \phi \partial_\nu \phi - \kappa m_p^2 \phi^{-\alpha} 
   \right) d^4x},
\end{equation}
where the parameter $\kappa$ is \citep{Peebles&Ratra1988,Anatoly2}
\begin{equation}
\kappa=\frac{8}{3}\left(\frac{2\alpha}{3}\right)^{\alpha /2}(\alpha +4)(\alpha +2)^{(\alpha -2)/2}.
\end{equation}

In this model, at the current
epoch, scalar field dark energy dominates the cosmological energy budget
and fuels the accelerating cosmological expansion. Prior to 
that space curvature dominated and at even earlier times non-relativistic
matter powered the decelerating cosmological expansion.
In the matter dominated epoch at
$a \ll a_0$, $\rho_\phi \ll \rho_{m}$ and $\rho_K \ll \rho_{m}$, the Einstein-de Sitter model applies, and the
initial conditions are that the cosmological scale factor evolves as $a(t)\propto t^{2/3}$,
the scalar field $\phi(t) \propto t^{2/(\alpha+2)}$, 
and the scalar field energy density evolves as 
$\rho_{\phi}\propto a^{-3\alpha/(\alpha+2)}\propto t^{-2\alpha/(\alpha+2)}$, as described in 
\cite{Peebles&Ratra1988}. In the space curvature dominated epoch $\rho_{\phi} \ll \rho_{K}$ and 
$\rho_{m} \ll \rho_{K}$ and $a(t) \propto t$, the scalar field $\phi (t) \propto t^{2/(2+\alpha)}$,
and the scalar field energy density evolves as $\rho_{\phi}\propto a^{-2\alpha/(2+\alpha)} \propto t^{-2\alpha/(2+\alpha)}$,\footnote[11]{As
long as the scalar field energy density does not dominate, the scalar field energy density $\rho_{\phi}\propto
t^{-2\alpha/(2+\alpha)}$, independent of the type of matter that dominates.} as determined in \cite{Anatoly2}. Hence, for positive
values of $\alpha$, the scalar field energy decreases, but less
rapidly than that of space curvature in the space curvature 
dominated epoch ($\rho_K \propto a^{-2}\propto t^{-2}$)
and less rapidly than that of non-relativistic matter in the matter dominated epoch 
($\rho_{m}\propto a^{-3} \propto t^{-2}$), so at late times 
the Universe will become dark energy dominated \citep{Anatoly2,Ratra&Peebles1988}. As in the radiation and matter
dominated epochs \citep{Peebles&Ratra1988,Ratra&Peebles1988}, \cite{Anatoly2} show that in the curvature 
dominated epoch the solution for the scalar field is a time-dependent fixed point or attractor. This means that for a wide 
range of initial conditions the solution will approach this special time-dependent fixed point solution. 

The equation of motion of the scalar field is, 
\begin{equation}
\label{eq:dotphi} 
    \ddot{\phi} + 3 \frac{\dot{a}}{a}\dot{\phi} -
    \frac{\kappa}{2} \alpha m_p^2 \phi^{-(\alpha+1)} = 0.
\end{equation}
In the presence of spatial curvature the $\phi$CDM model 
Friedmann equation takes the form
\begin{eqnarray}
\label{eq:phicdmfriedman}
   H^2(z; H_0, \textbf{p})\hspace{-2mm}
   &=&\hspace{-2mm} H_0^2[\Omega_{m0}(1+z)^3+\Omega_\phi (z,\alpha)+\Omega_{K0}(1+z)^2],
\end{eqnarray}
where the time-dependent scalar field density parameter $\Omega_{\phi}$ is defined as 
\begin{eqnarray}
\label{eq:Omegaphi} 
   \Omega_\phi(z,\alpha) \equiv \frac{8 \pi G}{3H^{2}_{0}}\rho_\phi= 
   \frac{1}{12H^{2}_{0}}\left(\dot\phi^2+\kappa m^{2}_{p}\phi^{-\alpha}\right).
\end{eqnarray}
In the limit $\alpha=0$ the $\phi$CDM model
is equivalent to the ordinary time-independent 
cosmological constant $\Lambda$ model. 
This makes the $\phi$CDM model a generalization of the 
standard $\Lambda$CDM model of cosmology.

Solving the coupled differential 
equations\ ({\ref{eq:dotphi}})---({\ref{eq:Omegaphi}}), 
with the initial conditions described in 
\cite{Peebles&Ratra1988} and \cite{Anatoly2}, allows for a numerical computation of 
the Hubble parameter $H(z; H_0, \textbf{p})$, as well
as the other functions needed for applications of the cosmological
tests. In this case the model parameters 
are taken to be $\textbf{p}=(\Omega_{m0},\alpha, \Omega_{K0})$.

\section{Observational constraints}
\label{sec:Constraints}

To constrain cosmological parameters \textbf{p}
we generalize the technique described in \cite{Farooq2013a} to models
with three free parameters, $\textbf{p}=(\Omega_{m0},\omega_X, \Omega_{K0})$ 
for the XCDM parameterization and $\textbf{p}=(\Omega_{m0},\alpha, \Omega_{K0})$ for
$\phi$CDM. Following \cite{Farooq2012} we compute a likelihood function 
$\mathcal{L}(\textbf{p})$ that depends on the three \textbf{p} parameters. We
compute these likelihood functions over the parameter ranges
$-0.7\leq \Omega_{K0} \leq 0.7$, $-2.0 \leq \omega_{X} \leq 0$, and  $ 0\leq \Omega_{m0} \leq 1.0$ 
for the XCDM parameterization, and $-0.2\leq \Omega_{K0} \leq 0.2$, $0 \leq \alpha \leq 5$, 
and  $ 0\leq \Omega_{m0} \leq 1.0$ for the $\phi$CDM model. For the sake of 
computational tractability the $\Omega_{K0}$ range considered in the case of $\phi$CDM
is much smaller than that used in the XCDM parameterization computation. 

To get two-dimensional likelihood functions $\mathcal{L}(\boldsymbol\theta)$, we
marginalize the three-dimensional likelihood function $\mathcal{L}(\textbf{p})$ over 
each of the three model parameters in turn, with flat priors.  Here
\begin{eqnarray}
\mathcal{L}(\boldsymbol\theta) \equiv \int \limits^{\beta_2}_{\beta_1}\mathcal{L}(\textbf{p})d\beta=\int \limits^{\beta_2}_{\beta_1}\mathcal{L}(\boldsymbol\theta,\beta)d\beta,
\label{eq:marg3to2}
\end{eqnarray}
where $\boldsymbol\theta$ is the set of two parameters at a time and $\beta$ is the
third parameter with marginalization limits of $\beta_1$ and $\beta_2$.

To maximize the two-dimensional likelihood function $\mathcal{L}(\boldsymbol\theta)$
we minimize $\chi^{2}(\boldsymbol\theta)\equiv
-2\mathrm{ln}\mathcal{L}(\boldsymbol\theta)$ with respect to model 
parameters $\boldsymbol\theta$ to find the best-fit 
parameter values $\boldsymbol\theta_0$. We define $1\sigma$, 
$2\sigma$, and $3\sigma$ confidence contours as two-dimensional 
parameter sets bounded by $\chi^2(\boldsymbol\theta) =
\chi^2(\boldsymbol\theta_0)+2.3,~\chi^2(\boldsymbol\theta) = 
\chi^2(\boldsymbol\theta_0)+6.17$, and $\chi^2(\boldsymbol\theta) = 
\chi^2(\boldsymbol\theta_0)+11.8$, respectively.\footnote[12]{While 
performing the data analysis described in \cite{farooq3} we found that the two-dimensional
contours obtained from integrating the likelihood function and
those obtained by using the $\chi^2$ prescription described here hardly differ. 
To save computational time we use the $\chi^2$ prescription in this paper.}

\subsection{Constraints from $H(z)$, SNIa, and BAO data sets, one at a time}

We first consider $H(z)$ data constraints. For this we use 22 independent $H(z)$ 
measurements and one standard deviation uncertainties at measured redshifts (covering the redshift range of 0.09 to 2.3),
listed in Table 1,\footnote[13]{In 
\cite{farooq3} we found that an augmented set of $H(z)$ measurements shows
clear evidence for the cosmological deceleration-acceleration transition predicted to 
occur in cosmological models dominated by dark energy at the current epoch. 
\cite{farooq4} more clearly illustrate the presence of this transition in the data by binning 
and combining the $H(z)$ data.} to constrain cosmological 
model parameters \textbf{p}. Using Eq.\ (18) of \cite{Farooq2012},
which is obtained after marginalizing over the nuisance 
parameter $H_0$ using a Gaussian prior with
$H_0 = 68 \pm 2.8$ km s$^{-1}$ Mpc$^{-1}$, \footnote[14]{As
discussed in \cite{Farooq2012}, the constraint contours are sensitive
to the $H_0$ prior used. The $H_0$ prior we use is obtained from a 
median statistics analysis \citep{Gott2001} of 553 $H_0$ measurements \citep{Chen2011a},
and has been stable now for more than a decade \citep{Gott2001,Chen2003}.
Recent measurements of $H_0$ are consistent with this value \citep[see e.g.,][]{colless12,ade13}
although some suggest slightly larger or smaller values \citep[see e.g., ][]{Freedman2012,
Sorce2012,Tammann2012}. It may be significant that the value of $H_0$ we use does
not demand the presence of dark radiation \citep{calabrese12}.}
we get a likelihood function $\mathcal{L}_H(\textbf{p})$ 
that depends only on model parameters \textbf{p} $=(\Omega_{m0}, 
\omega_X, \Omega_{K0})$ for the XCDM parameterization and $(\Omega_{m0}, \alpha, \Omega_{K0})$ for the 
$\phi$CDM model. Then using 
Eq.\ (\ref{eq:marg3to2}) we compute $\mathcal{L}_{H}(\boldsymbol\theta)$
from $\mathcal{L}_{H}(\textbf{p})$, and the two-dimensional confidence 
contours are obtained following the procedure discussed above.

\begin{center}
\begin{threeparttable}
\caption{Hubble Parameter Versus Redshift Data}
\vspace{5 mm}
\begin{tabular}{cccc}

\hline\hline
\multirow{2}{*}{$z$} & \multirow{1}{*}{$H(z)$} & \multirow{1}{*}{$\sigma_H$} & \multirow{2}{*}{Reference\tnote{a}} \\
                     & \multirow{1}{*}{(km s$^{-1}$ Mpc $^{-1}$)} & \multirow{1}{*}{(km s$^{-1}$ Mpc $^{-1}$)}  & \\
\hline
0.090&~~	69&~~	12&~~	1\\
0.170&~~	83&~~	8&~~	1\\
0.179&~~	75&~~	4&~~	4\\
0.199&~~	75&~~	5&~~	4\\
0.240&~~	79.69&~~	2.65&~~	2\\
0.270&~~	77&~~	14&~~	1\\
0.352&~~	83&~~	14&~~	4\\
0.400&~~	95&~~	17&~~	1\\
0.430&~~	86.45&~~	3.68&~~	2\\
0.480&~~	97&~~	62&~~	3\\
0.593&~~	104&~~	13&~~	4\\
0.680&~~	92&~~	8&~~	4\\
0.781&~~	105&~~	12&~~	4\\
0.875&~~	125&~~	17&~~	4\\
0.880&~~	90&~~	40&~~	3\\
0.900&~~	117&~~	23&~~	1\\
1.037&~~	154&~~	20&~~	4\\
1.300&~~	168&~~	17&~~	1\\
1.430&~~	177&~~	18&~~	1\\
1.530&~~	140&~~	14&~~	1\\
1.750&~~	202&~~	40&~~	1\\
2.300&~~	224&~~	8&~~	5\\
\hline
\end{tabular}
\begin{tablenotes}
\item[a]{(1) \cite{simon05}, (2) \cite{gaztanaga09}, (3) \cite{Stern2010}, 
(4) \cite{moresco12}, (5) \cite{Busca12}.}
\end{tablenotes}
\label{tab:Hz}
\end{threeparttable}
\end{center}


To tighten constraints on model parameters we also
use a second data
set,  the \cite{suzuki12} Union2.1 compilation of 580 SNIa distance 
modulus measurements at measured redshifts
(covering the redshift range of 0.015 to 1.414) with corresponding
one standard deviation uncertainties including systematic uncertainties. To constrain 
cosmological model parameters using this data the three-dimensional 
likelihood function $\mathcal{L}_{SN}(\textbf{p})$ is defined by
generalizing Eq.\ (26) of \cite{Farooq2012}, marginalizing over a 
flat $H_0$ prior (for these SNIa data). Then using 
Eq.\ (\ref{eq:marg3to2}) we determine $\mathcal{L}_{SN}(\boldsymbol\theta)$
from $\mathcal{L}_{SN}(\textbf{p})$, and the two-dimensional confidence 
contours are obtained as discussed above.
 
The third set of data we consider are the 6 BAO peak length scale
measurements (covering the redshift range of 0.1 to 0.75) with corresponding one standard deviation uncertainties, 
from \cite{Percival2010}, \cite{beutler2011}, and \cite{blake11}. 
To constrain model parameters \textbf{p}
we compute the three-dimensional likelihood function $\mathcal{L}_{BAO}(\textbf{p})$
by again marginalizing over a flat $H_0$ prior (for these BAO data), as discussed in
Sec.\ 5 of \cite{Farooq2012}. 
Then using Eq.\ (\ref{eq:marg3to2}) we compute $\mathcal{L}_{BAO}(\boldsymbol\theta)$
from $\mathcal{L}_{BAO}(\textbf{p})$, and the two-dimensional confidence 
contours are obtained using the procedure discussed above.

\begin{figure}[p]
\centering
  \includegraphics[width=48mm]{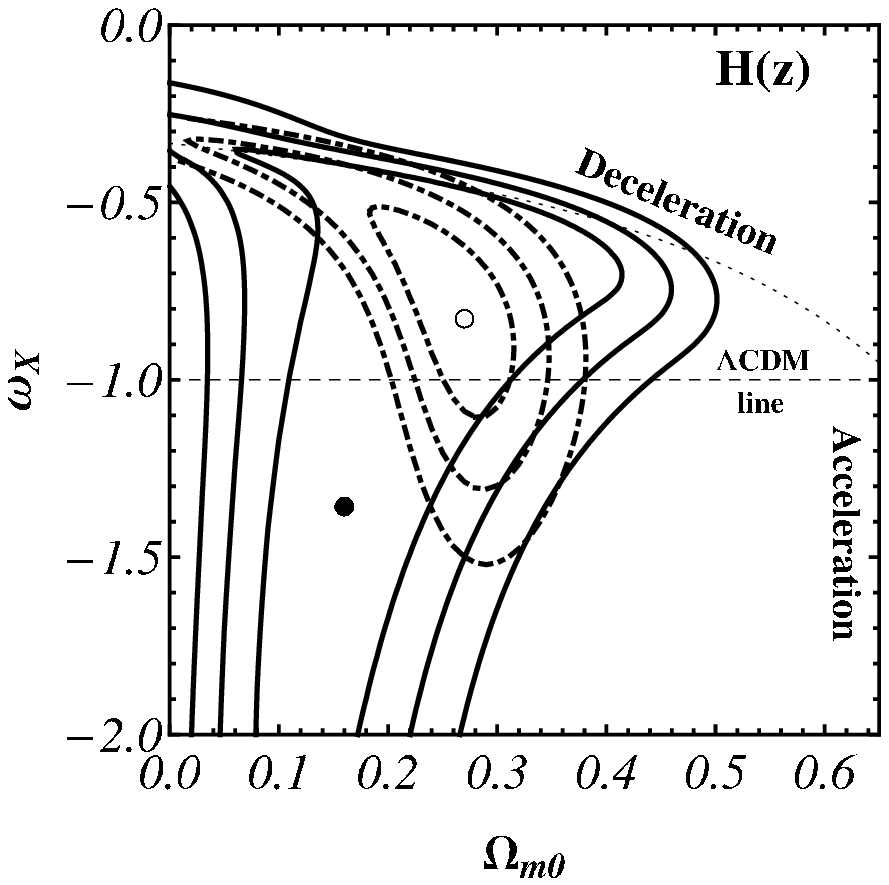}
  \includegraphics[width=50mm]{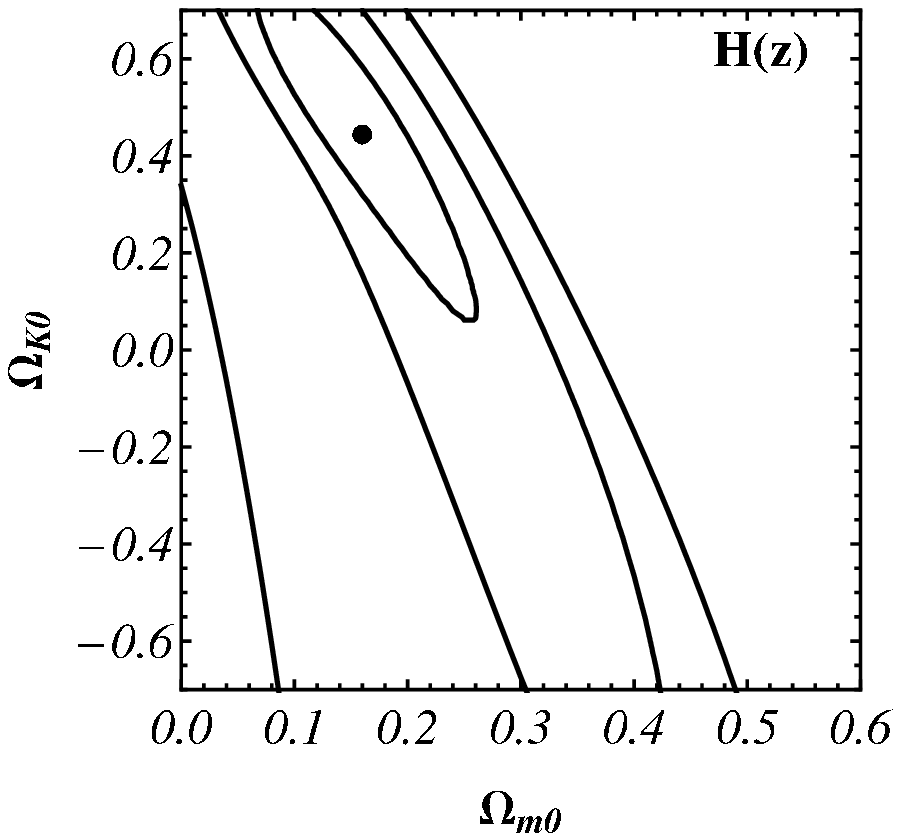}
  \includegraphics[width=50mm]{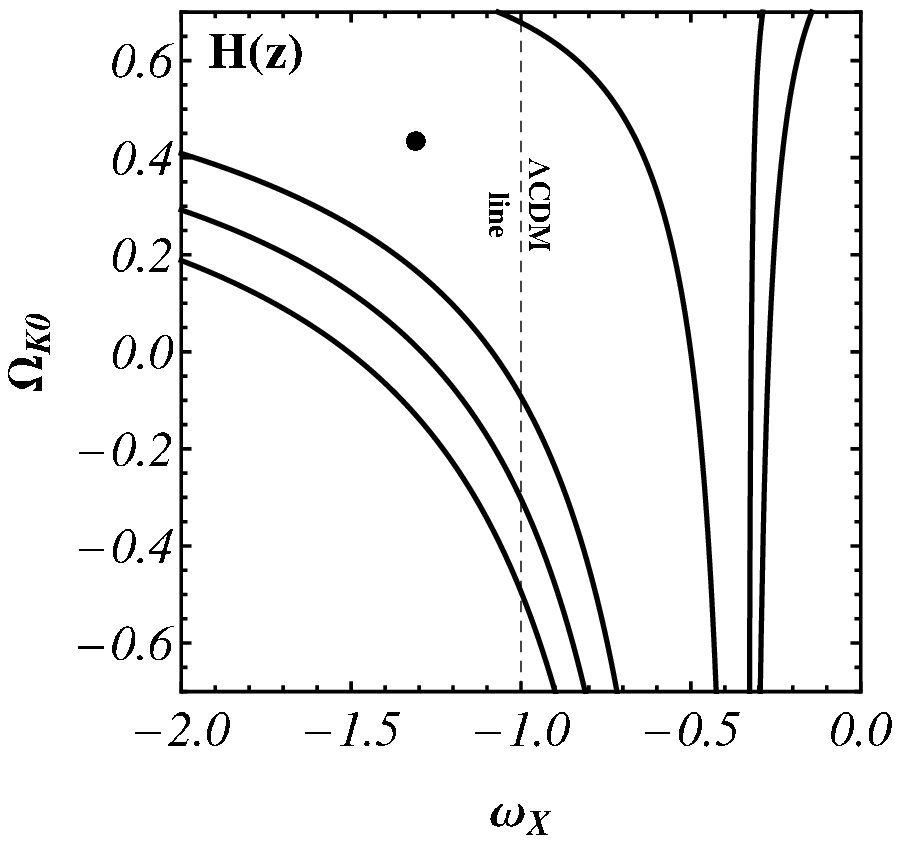}
  \includegraphics[width=48mm]{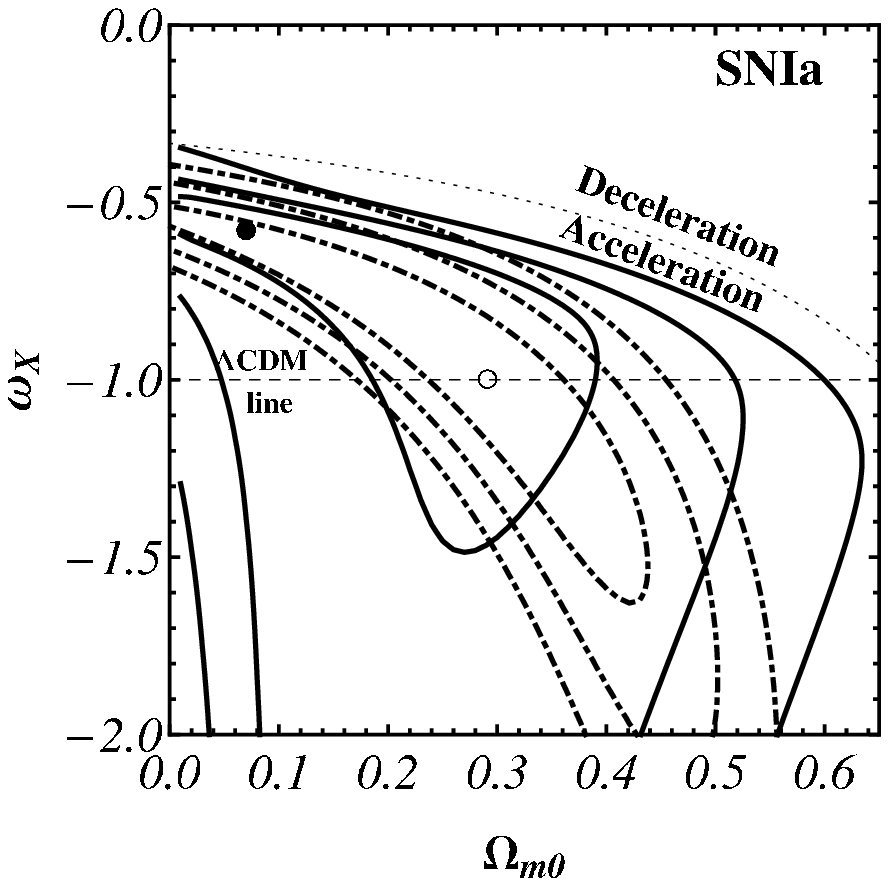}
  \includegraphics[width=50mm]{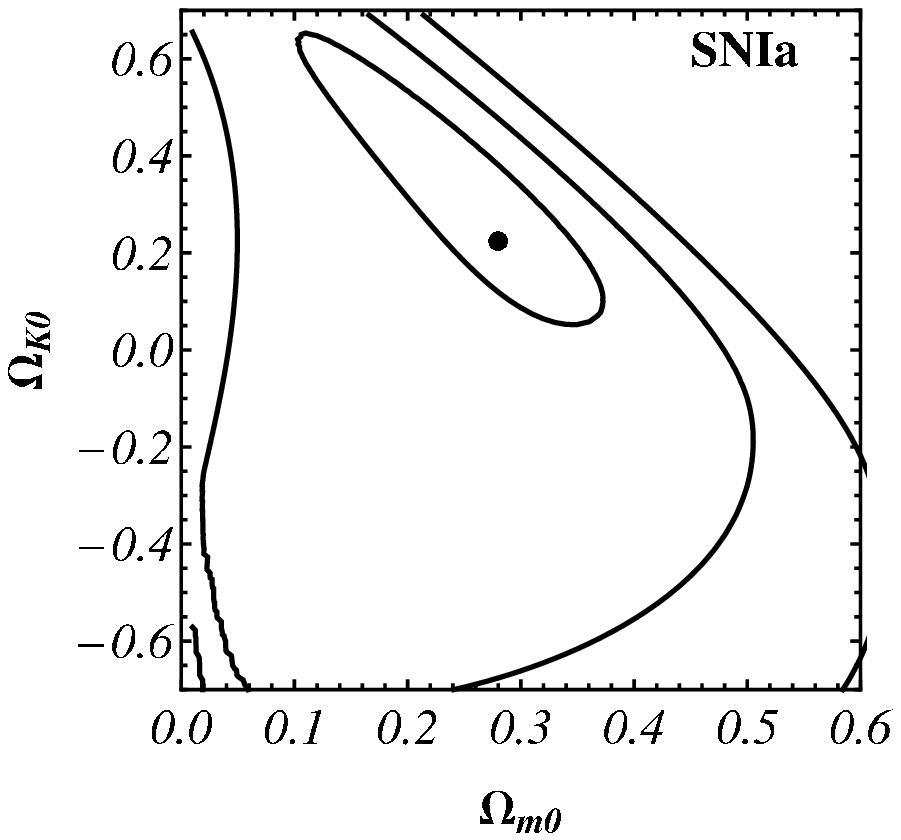}
  \includegraphics[width=50mm]{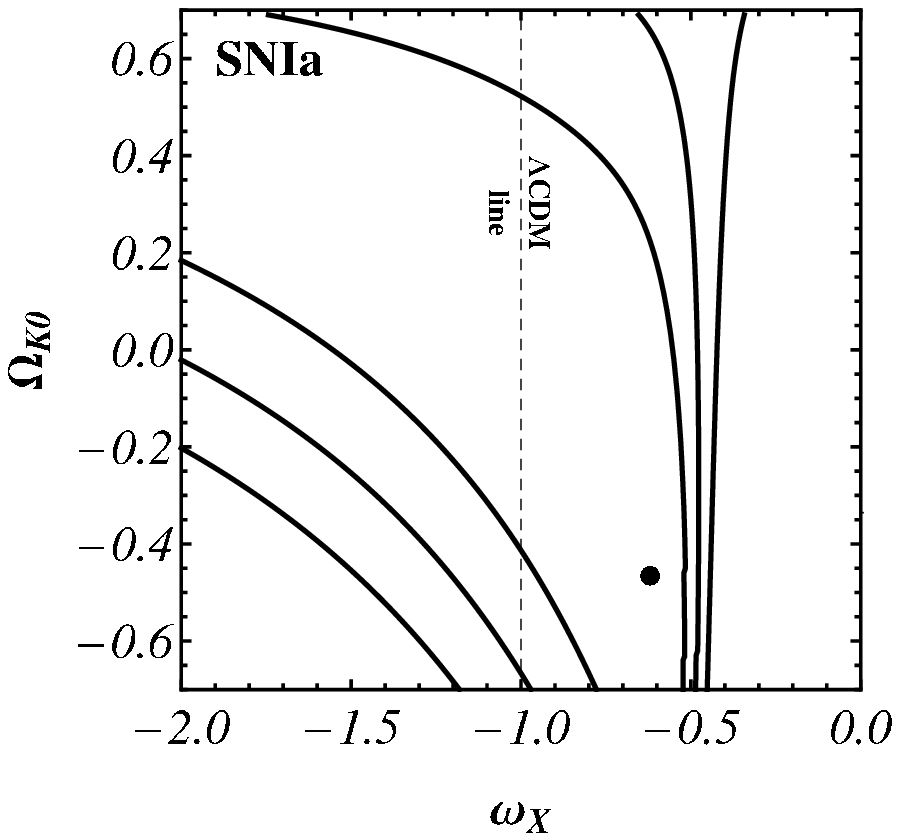}
  \includegraphics[width=48mm]{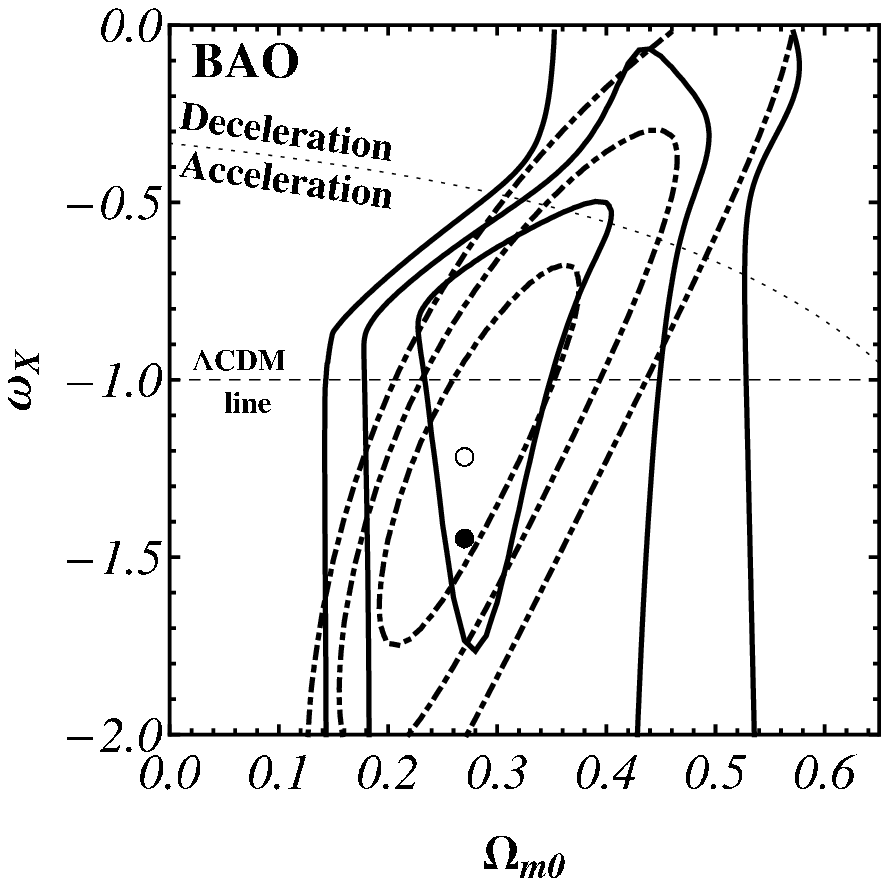}
  \includegraphics[width=50mm]{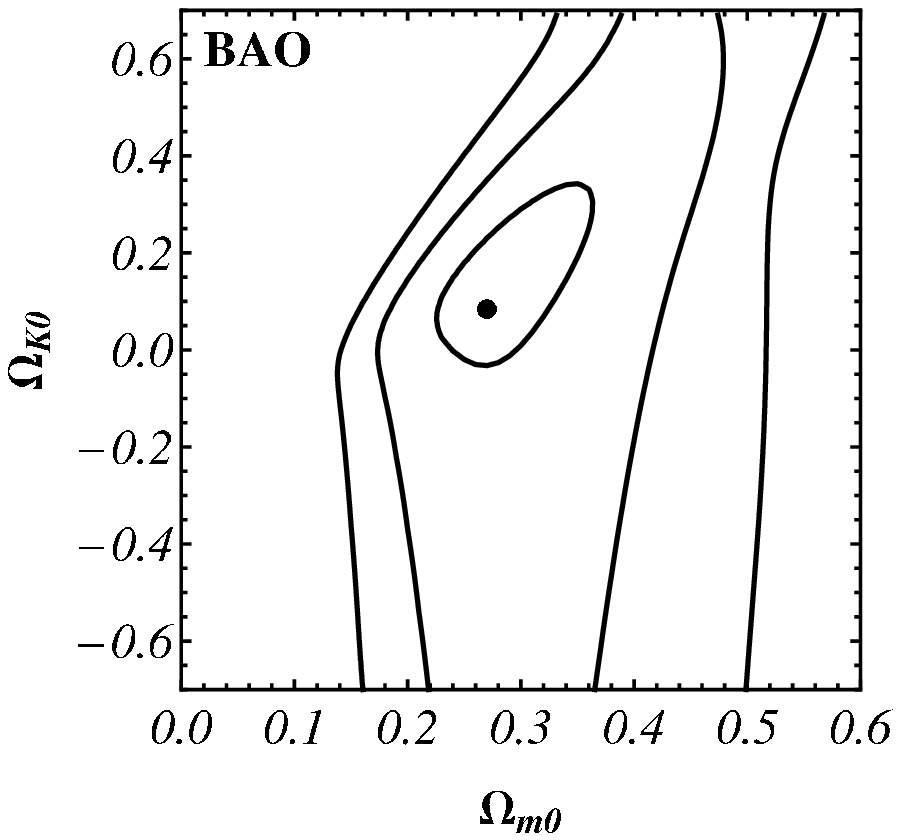}
  \includegraphics[width=50mm]{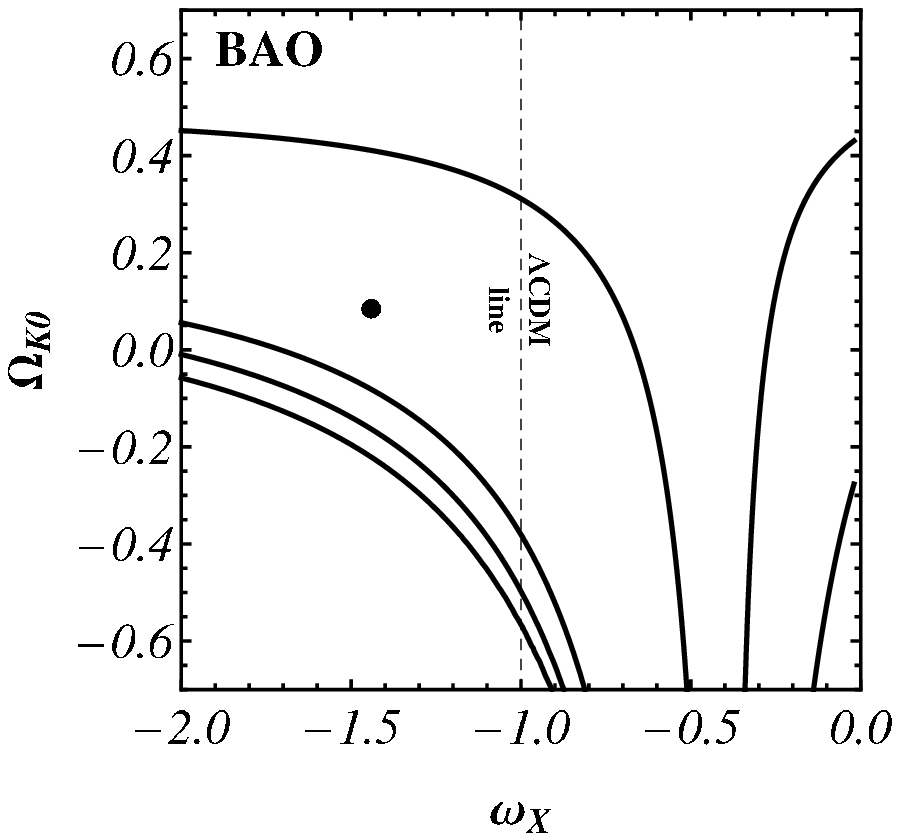}

\caption{
$1\sigma$, $2\sigma$, and $3\sigma$ constraint contours (solid lines) for parameters of the non-flat XCDM dark energy parameterization from 
$H(z)$ (first row), SNIa (second row), and BAO (third row) measurements; filled circles show best-fit points. The dot-dashed lines in the first column panels are $1\sigma$, $2\sigma$, and $3\sigma$ constraint contours
derived by \cite{Farooq2013a} using the spatially-flat XCDM parameterization (open circles show best-fit points); here dotted lines distinguish between
accelerating and decelerating models (at zero space curvature) and dashed lines (here and in the third column) correspond to the 
$\Lambda$CDM model. First, second, and third columns correspond to marginalizing over $\Omega_{K0}$, 
$\omega_X$, and $\Omega_{m0}$ respectively.
} \label{fig:XCDM_S}
\end{figure}

\begin{figure}[p]
\centering
  \includegraphics[width=46mm]{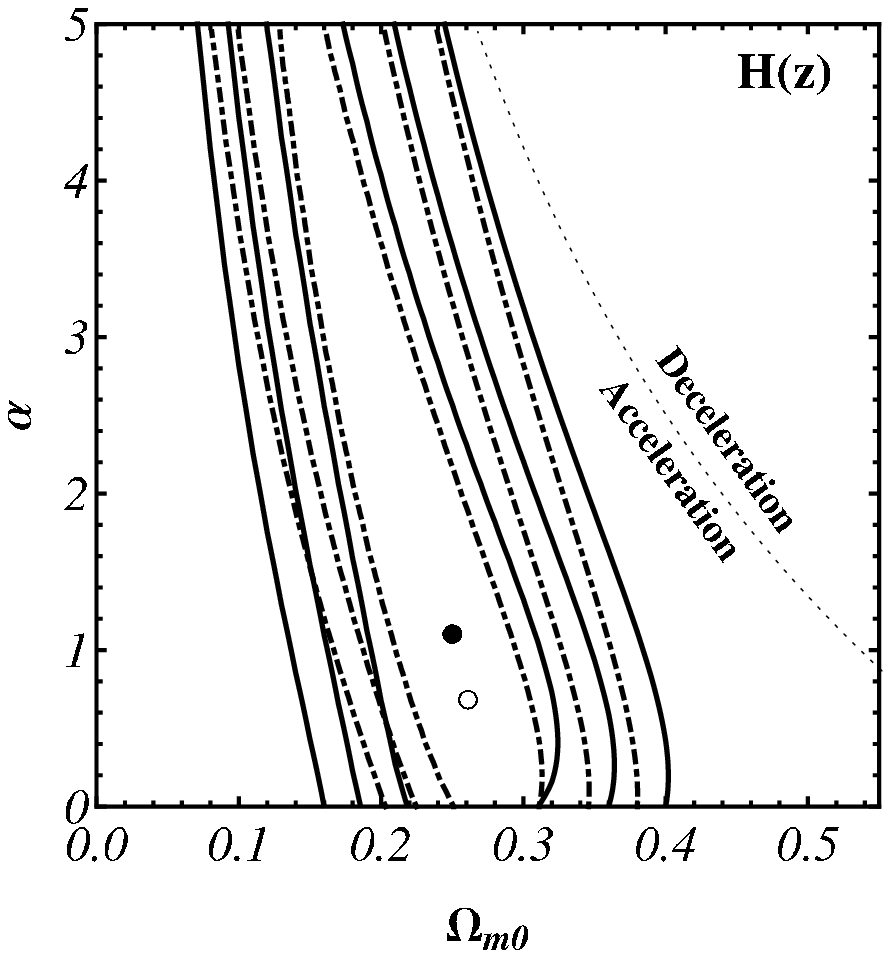}
  \includegraphics[width=52mm]{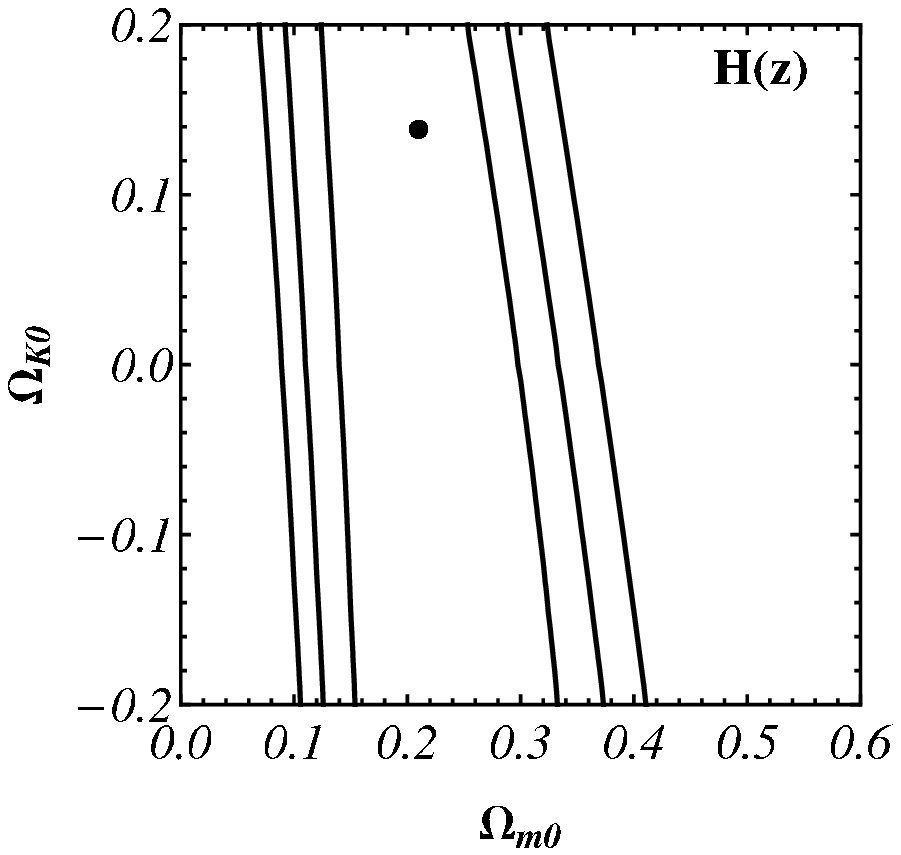}
  \includegraphics[width=52mm]{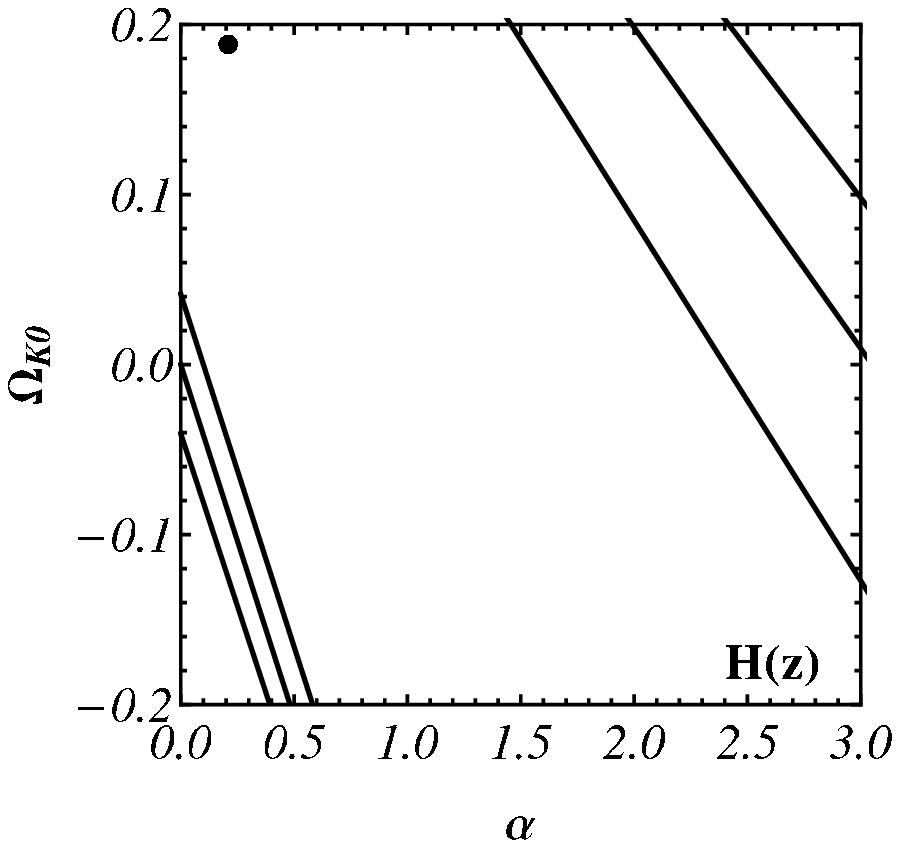}
  \includegraphics[width=46mm]{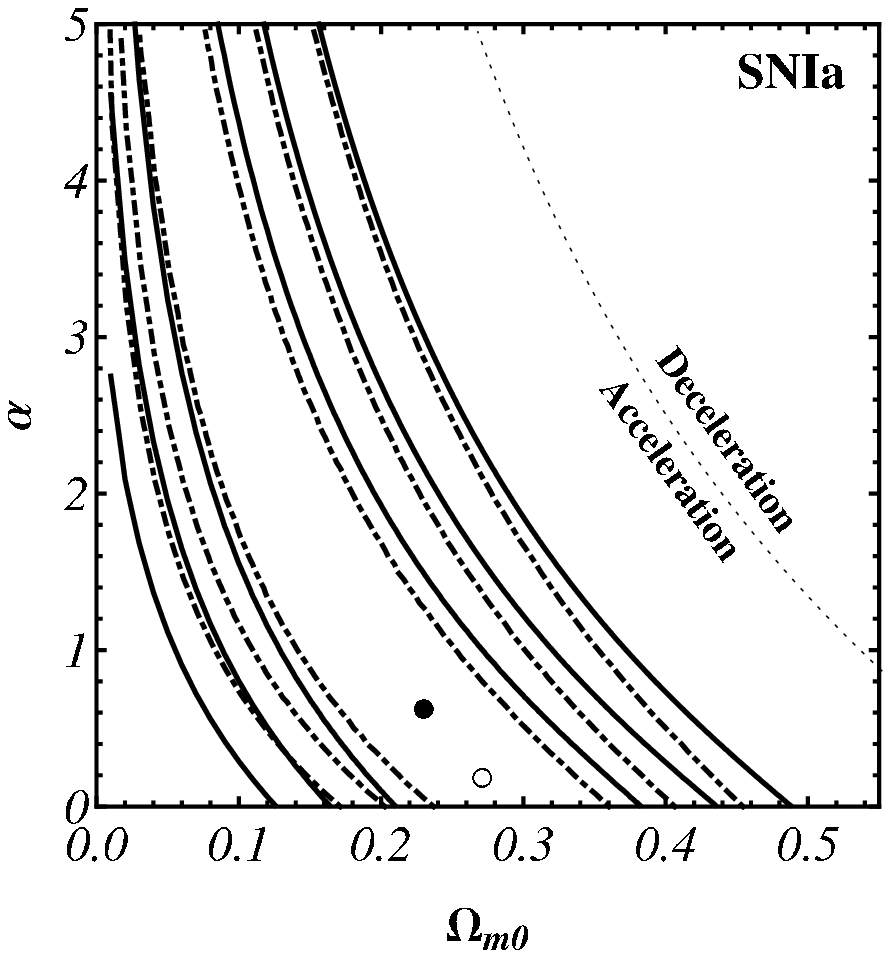}
  \includegraphics[width=52mm]{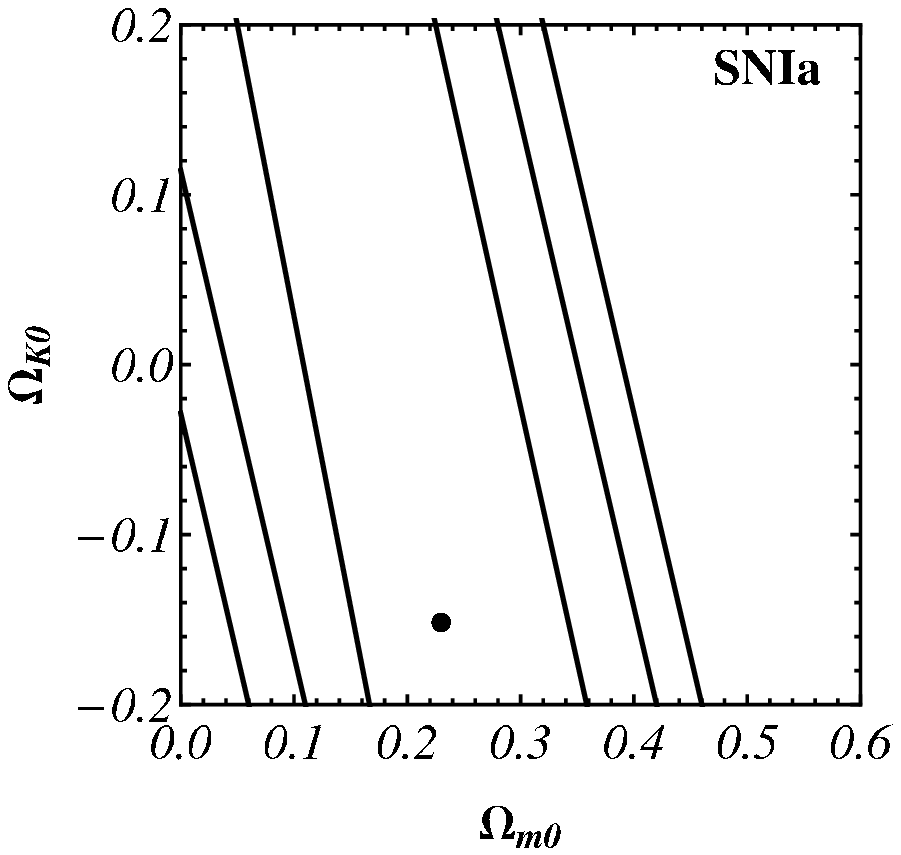}
  \includegraphics[width=52mm]{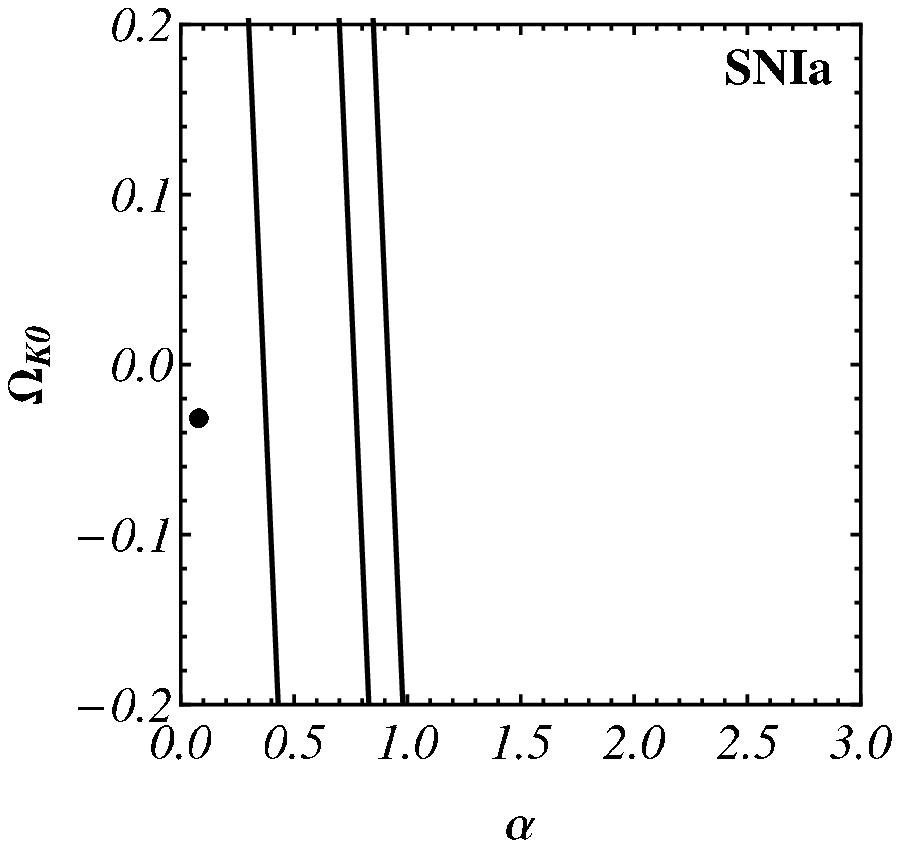}
  \includegraphics[width=46mm]{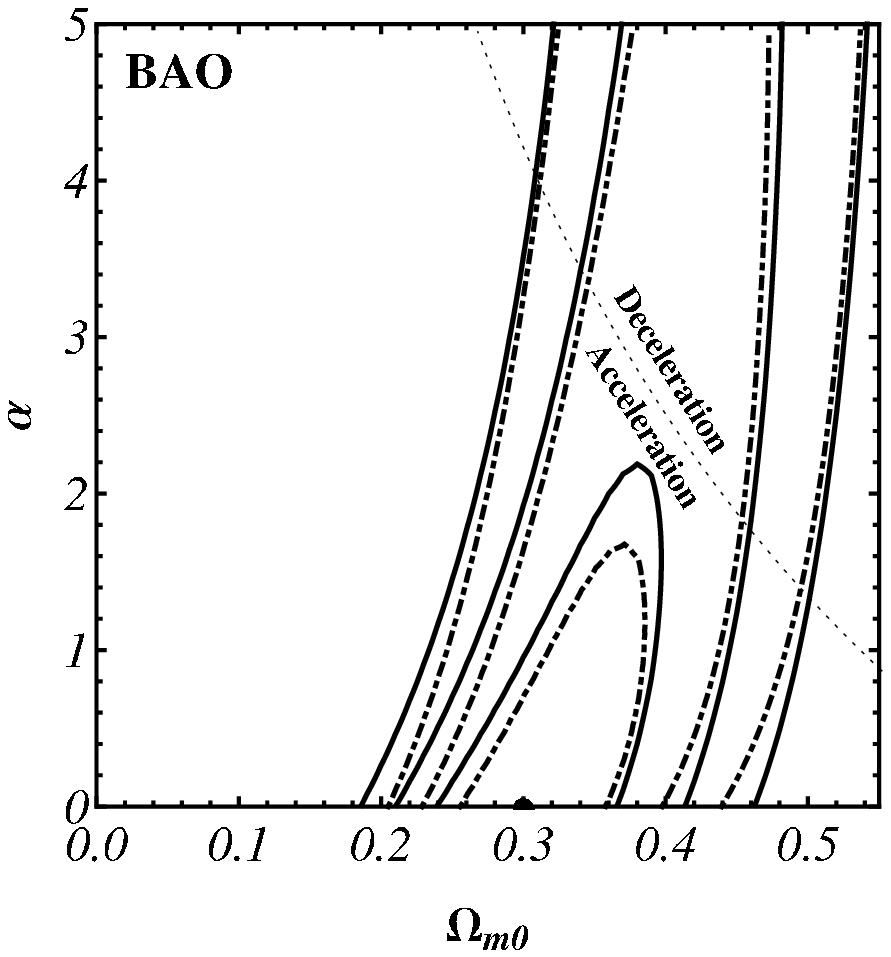}
  \includegraphics[width=52mm]{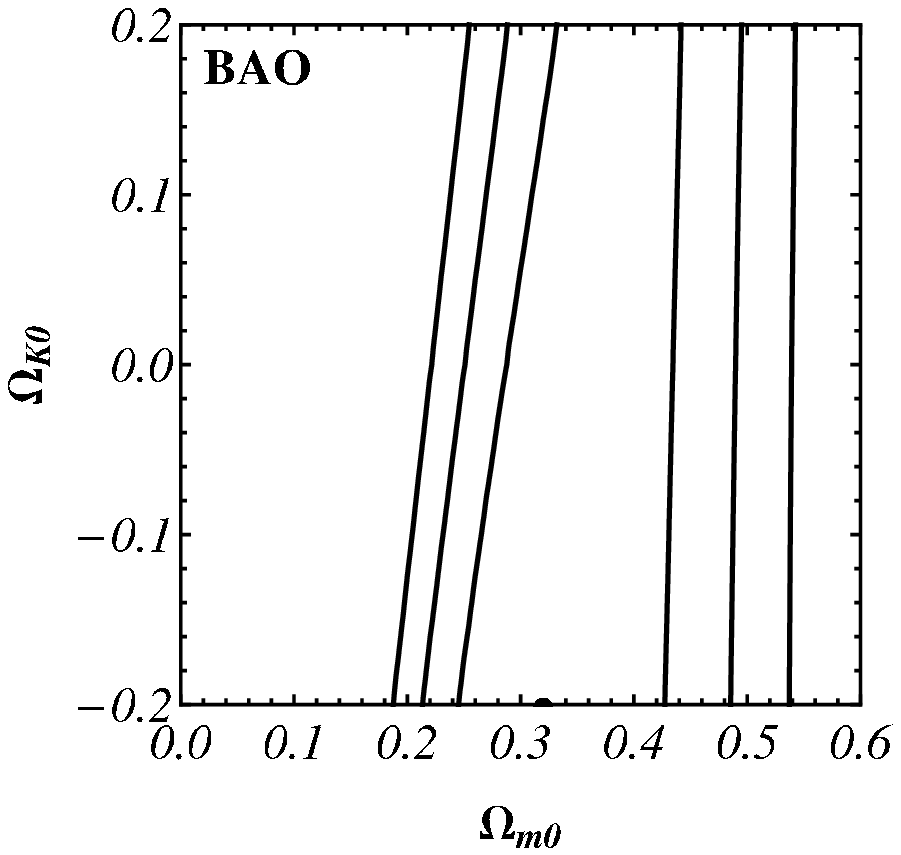}
  \includegraphics[width=52mm]{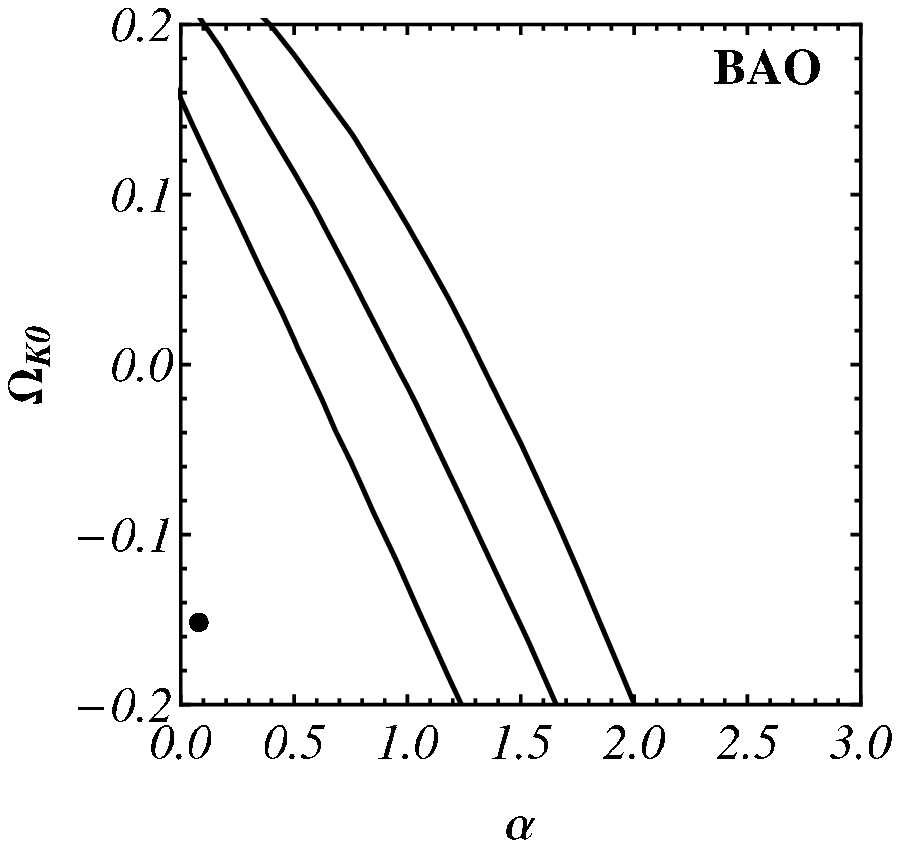}
\caption{
$1\sigma$, $2\sigma$, and $3\sigma$ constraint contours (solid lines) for parameters of the non-flat $\phi$CDM dark energy model from 
$H(z)$ (first row), SNIa (second row), and BAO (third row) measurements; filled circles show best-fit points. The dot-dashed lines in the first column panels are $1\sigma$, $2\sigma$, and $3\sigma$ constraint contours
derived by \cite{Farooq2013a} using the spatially-flat $\phi$CDM model (open circles show best-fit points); here dotted lines distinguish between
accelerating and decelerating models (at zero space curvature) and  the $\alpha=0$ axes (here and in the third column) correspond to the $\Lambda$CDM model. First, second, and third columns correspond to marginalizing over $\Omega_{K0}$, $\alpha$, and $\Omega_{m0}$ respectively. {We note that the $\Omega_{K0}=0$ constraints and the marginalized $\Omega_{K0}$ constraints differ by only a small amount due to the prior range of $\Omega_{K0}$ used here.} 
} \label{fig:phiCDM_S}
\end{figure}


\begin{center}
\begin{threeparttable}
\caption{XCDM Parameterization Results}
\begin{tabular}{ccccc}

\hline\hline
\multirow{1}{*}{Data Set} &  Marginalization Range &  Best-Fit Point& $\chi^2_{\mathrm{min}}$ & $\chi^2_{\mathrm{min}}/\mathrm{d.o.f}$ \\
\hline
\multirow{4}{*}{$H(z)$}& $ \Omega_{K0}=0$\tnote{a} & $(\Omega_{m0}, \omega_X)=(0.27,-0.82)$ & 15.2&0.800\\
\cline{2-5} 
{}& $-0.7\leqslant \Omega_{K0} \leqslant 0.7$ & $(\Omega_{m0}, \omega_X)=(0.16,-1.35)$ & 17.8&0.937\\ 
{}& $-2\leqslant \omega_{X} \leqslant 0$ & $(\Omega_{m0}, \Omega_{K0})=(0.16,0.45)$ & 14.5&0.763\\
{}& $0\leqslant \Omega_{m0} \leqslant 1$ & $(\omega_X, \Omega_{K0})=(-1.31,0.44)$ & 20.4&1.074\\
\hline
\multirow{4}{*}{$\mathrm{SNIa}$} & $ \Omega_{K0}=0$\tnote{a} & $(\Omega_{m0}, \omega_X)=(0.29,-0.99)$ & 545&0.945\\
\cline{2-5} 
{} & $-0.7\leqslant \Omega_{K0} \leqslant 0.7$ & $(\Omega_{m0}, \omega_X)=(0.07,-0.57)$ & 546&0.946\\
{} & $-2\leqslant \omega_{X} \leqslant 0$ & $(\Omega_{m0}, \Omega_{K0})=(0.28,0.23)$ & 545&0.945\\
{} & $0\leqslant \Omega_{m0} \leqslant 1$ & $(\omega_X, \Omega_{K0})=(-0.62,-0.46)$ & 549&0.951\\
\hline
\multirow{4}{*}{$\mathrm{BAO}$} & $ \Omega_{K0}=0$\tnote{a} & $(\Omega_{m0}, \omega_X)=(0.27,-1.21)$ & 5.50&1.833\\
\cline{2-5} 
{} & $-0.7\leqslant \Omega_{K0} \leqslant 0.7$ & $(\Omega_{m0}, \omega_X)=(0.27,-1.44)$ & 6.50&2.167\\
{} & $-2\leqslant \omega_{X} \leqslant 0$ & $(\Omega_{m0}, \Omega_{K0})=(0.27,0.09)$ & 4.90&1.633\\
{} & $0\leqslant \Omega_{m0} \leqslant 1$ & $(\omega_X, \Omega_{K0})=(-1.44,-0.09)$ & 10.4&3.467\\
\hline
\multirow{4}{*}{$H(z)+\mathrm{SNIa}$} & $ \Omega_{K0}=0$\tnote{a} & $(\Omega_{m0}, \omega_X)=(0.27,-0.90)$ & 561&0.937\\
\cline{2-5}
{} & $-0.7\leqslant \Omega_{K0} \leqslant 0.7$ & $(\Omega_{m0}, \omega_X)=(0.24,-0.97)$ & 562&0.938\\
{} & $-2\leqslant \omega_{X} \leqslant 0$ & $(\Omega_{m0}, \Omega_{K0})=(0.18,0.41)$ & 561&0.937\\
{} & $0\leqslant \Omega_{m0} \leqslant 1$ & $(\omega_X, \Omega_{K0})=(-0.98,0.15)$ & 566&0.945\\
\hline
\multirow{4}{*}{$H(z)+\mathrm{BAO}$} & $ \Omega_{K0}=0$\tnote{a} & $(\Omega_{m0}, \omega_X)=(0.29,-0.99)$ & 22.4&0.896\\
\cline{2-5}
{} & $-0.7\leqslant \Omega_{K0} \leqslant 0.7$ & $(\Omega_{m0}, \omega_X)=(0.31,-0.79)$ & 24.2&0.968\\
{} & $-2\leqslant \omega_{X} \leqslant 0$ & $(\Omega_{m0}, \Omega_{K0})=(0.31,-0.19)$ & 26.9&1.076\\
{} & $0\leqslant \Omega_{m0} \leqslant 1$ & $(\omega_X, \Omega_{K0})=(-0.78,-0.19)$ & 27.5&1.100\\
\hline
\multirow{4}{*}{$\mathrm{SNIa}+\mathrm{BAO}$} & $ \Omega_{K0}=0$\tnote{a} & $(\Omega_{m0}, \omega_X)=(0.30,-1.03)$ & 551&0.945\\
\cline{2-5}
{} & $-0.7\leqslant \Omega_{K0} \leqslant 0.7$ & $(\Omega_{m0}, \omega_X)=(0.29,-0.77)$ & 553&0.949\\
{} & $-2\leqslant \omega_{X} \leqslant 0$ & $(\Omega_{m0}, \Omega_{K0})=(0.31,0.22)$ & 552&0.947\\
{} & $0\leqslant \Omega_{m0} \leqslant 1$ & $(\omega_X, \Omega_{K0})=(-0.93,-0.10)$ & 556&0.954\\
\hline
\multirow{4}{*}{$H(z)+\mathrm{SNIa}+\mathrm{BAO}$} & $ \Omega_{K0}=0$\tnote{a} & $(\Omega_{m0}, \omega_X)=(0.31,-1.02)$ & 566&0.936\\
\cline{2-5}
{}& $-0.7\leqslant \Omega_{K0} \leqslant 0.7$ & $(\Omega_{m0}, \omega_X)=(0.30,-0.88)$ & 571&0.944\\
{} & $-2\leqslant \omega_{X} \leqslant 0$ & $(\Omega_{m0}, \Omega_{K0})=(0.29,-0.15)$ & 582&0.962\\
{} & $0\leqslant \Omega_{m0} \leqslant 1$ & $(\omega_X, \Omega_{K0})=(-0.90,-0.10)$ & 573&0.947\\
\hline
\hline
\end{tabular}
\begin{tablenotes}
\item[a]{From \cite{Farooq2013a}.}
\end{tablenotes}
\label{table:XCDMresults}
\end{threeparttable}
\end{center}



\begin{center}
\begin{threeparttable}
\caption{$\phi$CDM Model Results}
\begin{tabular}{ccccc}

\hline\hline
\multirow{1}{*}{Data Set} &  Marginalization Range &  Best-Fit Point& $\chi^2_{\mathrm{min}}$ & $\chi^2_{\mathrm{min}}/\mathrm{d.o.f}$ \\
\hline
\multirow{4}{*}{$H(z)$} & $ \Omega_{K0} =0$\tnote{a} & $(\Omega_{m0}, \alpha)=(0.26,0.70)$ & 15.2 & 0.800\\
\cline{2-5}
{} & $-0.2\leqslant \Omega_{K0} \leqslant 0.2$ & $(\Omega_{m0}, \alpha)=(0.25,1.12)$ & 17.8& 0.937\\
{} & $0\leqslant \alpha \leqslant 5$ & $(\Omega_{m0}, \Omega_{K0})=(0.21,0.14)$ & 13.9&0.732\\
{} & $0\leqslant \Omega_{m0} \leqslant 1$ & $(\alpha, \Omega_{K0})=(0.21,0.19)$ & 20.4&1.074\\
\hline
\multirow{4}{*}{$\mathrm{SNIa}$} & $ \Omega_{K0} =0$\tnote{a} & $(\Omega_{m0}, \alpha)=(0.27,0.20)$ & 545&0.945\\
\cline{2-5}
{} & $-0.2\leqslant \Omega_{K0} \leqslant 0.2$ & $(\Omega_{m0}, \alpha)=(0.23,0.64)$ & 548&0.948\\
{} & $0\leqslant \alpha \leqslant 5$ & $(\Omega_{m0}, \Omega_{K0})=(0.23,-0.15)$ & 547&0.948\\
{} & $0\leqslant \Omega_{m0} \leqslant 1$ & $(\alpha, \Omega_{K0})=(0.08,-0.03)$ & 550&0.953\\
\hline
\multirow{4}{*}{$\mathrm{BAO}$} & $ \Omega_{K0} =0$\tnote{a} & $(\Omega_{m0}, \alpha)=(0.30,0.00)$ & 5.9&1.967\\
\cline{2-5}
{} & $-0.2\leqslant \Omega_{K0} \leqslant 0.2$ & $(\Omega_{m0}, \alpha)=(0.30,0.01)$ & 8.30&2.767\\
{} &  $0\leqslant \alpha \leqslant 5$ & $(\Omega_{m0}, \Omega_{K0})=(0.32,-0.20)$ & 5.50&1.833\\
{} & $0\leqslant \Omega_{m0} \leqslant 1$ & $(\alpha, \Omega_{K0})=(0.08,-0.15)$ & 10.6&3.533\\
\hline
\multirow{4}{*}{$H(z)+\mathrm{SNIa}$} & $ \Omega_{K0} =0$\tnote{a} & $(\Omega_{m0}, \alpha)=(0.26,0.35)$ & 561&0.937\\
\cline{2-5}
{} & $-0.2\leqslant \Omega_{K0} \leqslant 0.2$ & $(\Omega_{m0}, \alpha)=(0.26,0.31)$ & 564&0.942\\
{} & $0\leqslant \alpha \leqslant 5$ & $(\Omega_{m0}, \Omega_{K0})=(0.25,0.11)$ & 562&0.938\\
{} & $0\leqslant \Omega_{m0} \leqslant 1$ & $(\alpha, \Omega_{K0})=(0.09,0.08)$ & 567&0.947\\
\hline
\multirow{4}{*}{$H(z)+\mathrm{BAO}$} & $ \Omega_{K0} =0$\tnote{a} & $(\Omega_{m0}, \alpha)=(0.29,0.00)$ & 22.4&0.896\\
\cline{2-5}
{} & $-0.2\leqslant \Omega_{K0} \leqslant 0.2$ & $(\Omega_{m0}, \alpha)=(0.30,0.34)$ & 25.2&1.008\\
{} & $0\leqslant \alpha \leqslant 5$ & $(\Omega_{m0}, \Omega_{K0})=(0.31,-0.20)$ & 21.9&0.876\\
{} & $0\leqslant \Omega_{m0} \leqslant 1$ & $(\alpha, \Omega_{K0})=(0.77,-0.20)$ & 27.5&1.100\\
\hline
\multirow{4}{*}{$\mathrm{SNIa}+\mathrm{BAO}$} & $ \Omega_{K0} =0$\tnote{a} & $(\Omega_{m0}, \alpha)=(0.30,0.00)$ & 551&0.945\\
\cline{2-5}
{} & $-0.2\leqslant \Omega_{K0} \leqslant 0.2$ & $(\Omega_{m0}, \alpha)=(0.30,0.08)$ & 554&0.950\\
{} & $0\leqslant \alpha \leqslant 5$ & $(\Omega_{m0}, \Omega_{K0})=(0.30,-0.05)$ & 553&0.949\\
{} & $0\leqslant \Omega_{m0} \leqslant 1$ & $(\alpha, \Omega_{K0})=(0.02,-0.03)$ & 557&0.955\\
\hline
\multirow{4}{*}{$H(z)+\mathrm{SNIa}+\mathrm{BAO}$} & $ \Omega_{K0} =0$\tnote{a} & $(\Omega_{m0}, \alpha)=(0.29,0.00)$ & 567&0.937\\
\cline{2-5}
{} & $-0.2\leqslant \Omega_{K0} \leqslant 0.2$ & $(\Omega_{m0}, \alpha)=(0.30,0.46)$ & 571&0.944\\
{} & $0\leqslant \alpha \leqslant 5$ & $(\Omega_{m0}, \Omega_{K0})=(0.30,-0.05)$ & 569&0.940\\
{} & $0\leqslant \Omega_{m0} \leqslant 1$ & $(\alpha, \Omega_{K0})=(0.01,0.00)$ & 573&0.947\\
\hline
\hline
\end{tabular}
\begin{tablenotes}
\item[a]{From \cite{Farooq2013a}.}
\end{tablenotes}
\label{table:phiCDMresults}

\end{threeparttable}
\end{center}


Figures\ \ref{fig:XCDM_S} and
\ref{fig:phiCDM_S} show the constraints on parameters of the XCDM parameterization
and the $\phi$CDM model from the $H(z)$ (top row), SNIa (middle row), 
and BAO (bottom row) measurements. In these figures the panels in the first, second, and
third columns show the two-dimensional probability density 
constraint contours (solid lines) from $\mathcal{L}(\Omega_{m0},\omega_X)
[\mathcal{L}(\Omega_{m0},\alpha)]$, $\mathcal{L}(\Omega_{m0},\Omega_{K0})$, and
$\mathcal{L}(\omega_X,\Omega_{K0})[\mathcal{L}(\alpha,\Omega_{K0})]$ for the
XCDM parameterization [the $\phi$CDM model]. The dot-dashed contours in the panels of
the first columns of Figs.\ (\ref{fig:XCDM_S}) and 
(\ref{fig:phiCDM_S}) are $1\sigma$, $2\sigma$, and $3\sigma$ confidence contours
corresponding to spatially-flat models, reproduced from \cite{Farooq2013a}.
Tables\ \ref{table:XCDMresults}
and \ref{table:phiCDMresults} list best-fit points and $\chi^2_{\mathrm{min}}$ values.

Comparing the solid contours to the dot-dashed contours in the panels in the first columns of 
Figs.\ \ref{fig:XCDM_S} and \ref{fig:phiCDM_S}, we see that the addition of space curvature
as a third free parameter results in a fairly significant broadening of the 
constraint contours, as might have been anticipated. For the XCDM parameterization (first column of Fig.\ \ref{fig:XCDM_S}), since
the data constrain $\omega_X$ reasonably well in the spatially-flat case, the
inclusion of space curvature as a free parameter significantly weakens the bounds on
$\omega_X$. For the $\phi$CDM model (first column of Fig.\ \ref{fig:phiCDM_S})
the data do not constrain $\alpha$ (the corresponding parameter that governs
the time-variability of dark energy in this case) as tightly in the 
spatially-flat case, so inclusion of space curvature appears to have a relatively
less significant effect (this is probably also a consequence of the significantly smaller
$\Omega_{k0}$ range considered, $-0.2 \leq \Omega_{K0} \leq 0.2$, for
computational tractability). This interplay between space curvature and the parameter that 
governs the time-variability of dark energy is also evident in the second
and third columns of panels of Figs.\ \ref{fig:XCDM_S} and \ref{fig:phiCDM_S}.
Clearly, for the single data sets, including space curvature in the analysis
significantly weakens the support for a constant cosmological constant $\Lambda$,
while allowing dark energy density to be dynamical significantly weakens 
support for a spatially-flat model. The reason for this is that the space 
curvature energy density redshifts in a way that is closer to the behavior 
of the dark energy density than does the non-relativistic matter density.

These results show very clearly that when spatial curvature is a free parameter a 
single data set cannot significantly constrain cosmological parameters of
the dynamical dark energy models considered here. 
To tighten constraints on cosmological parameters, we next consider combinations of data sets.

\subsection{Constraints from combinations of data sets}

\begin{figure}[p]
\centering
  \includegraphics[width=50mm]{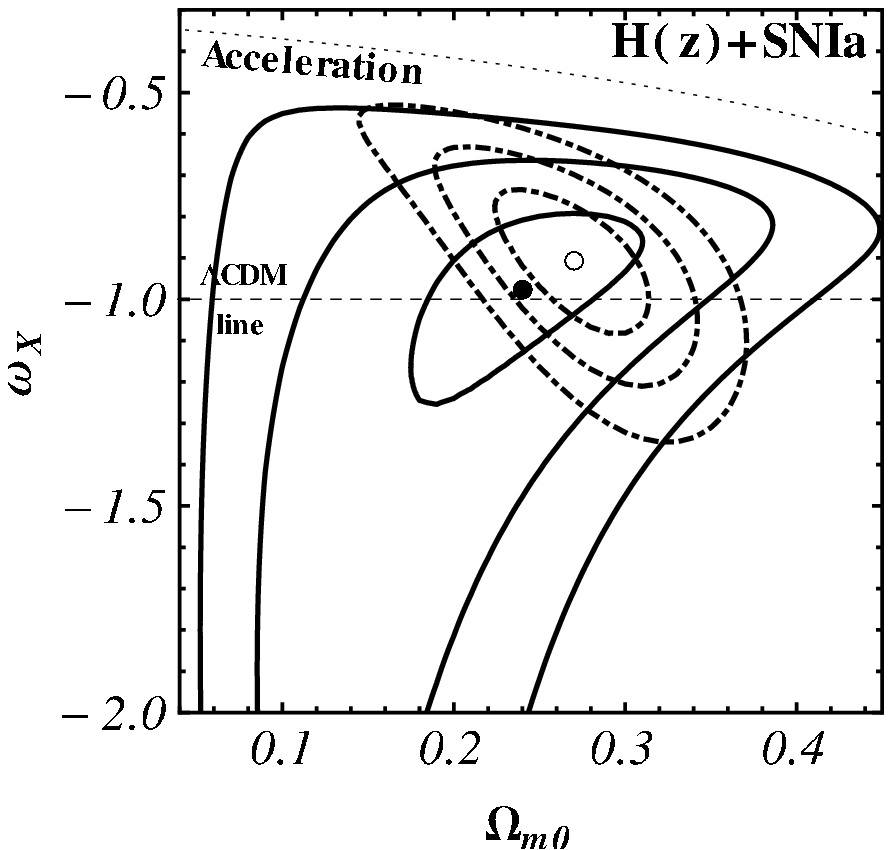}
  \includegraphics[width=50mm]{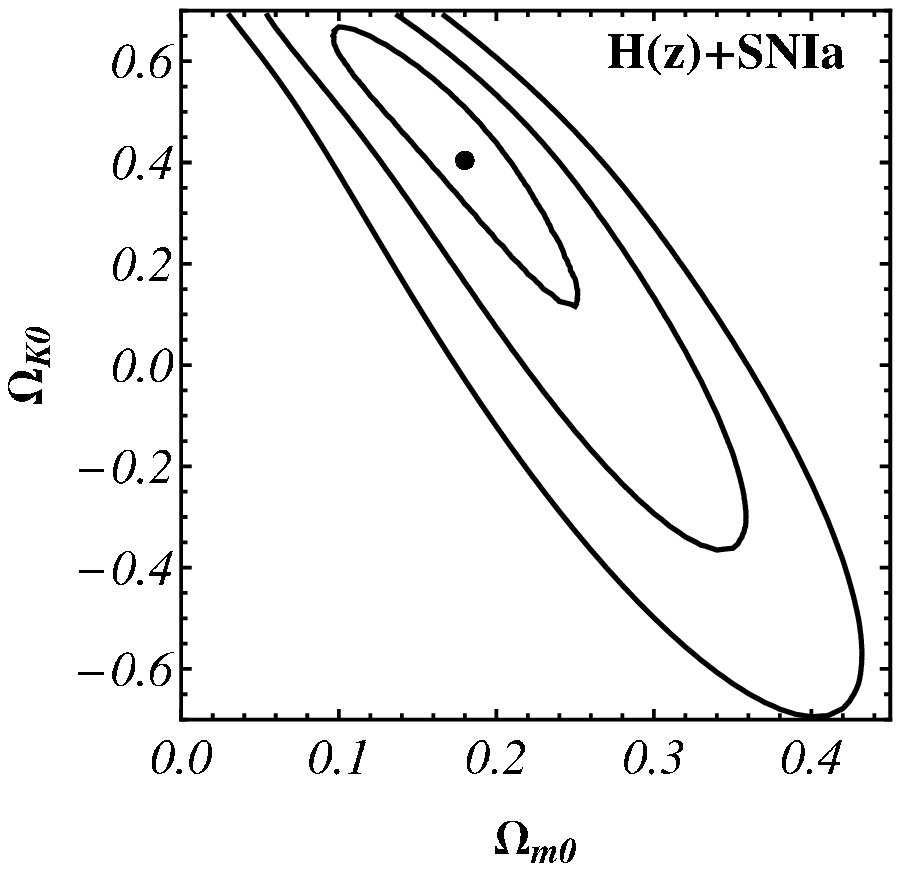}
  \includegraphics[width=50mm]{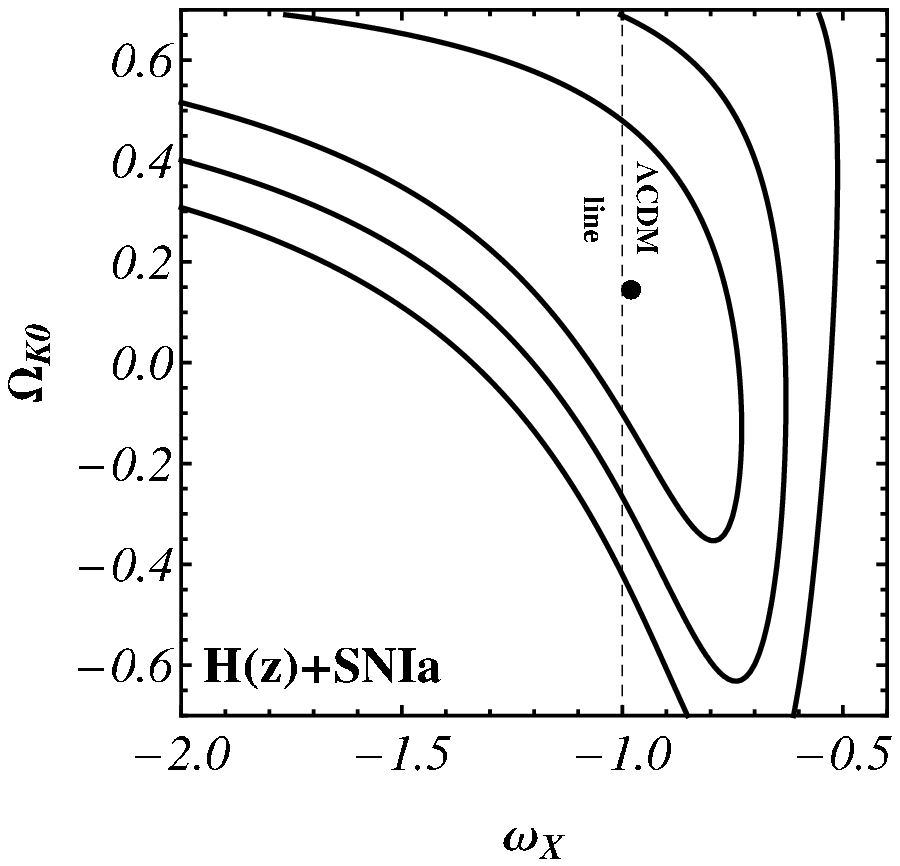}
  \includegraphics[width=50mm]{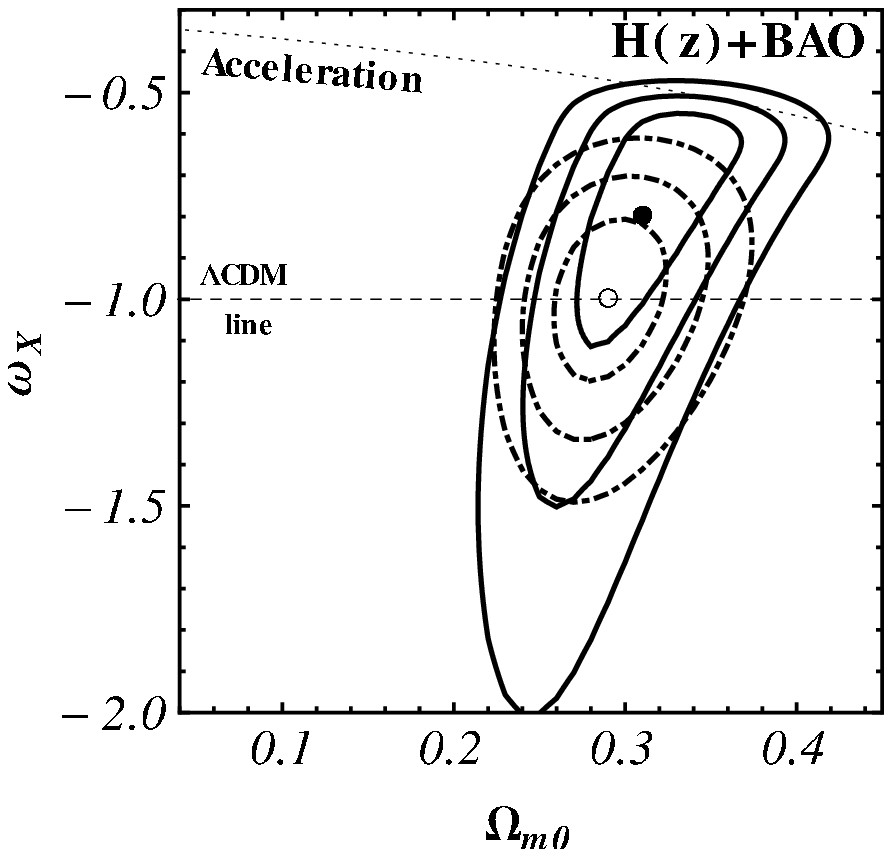}
  \includegraphics[width=50mm]{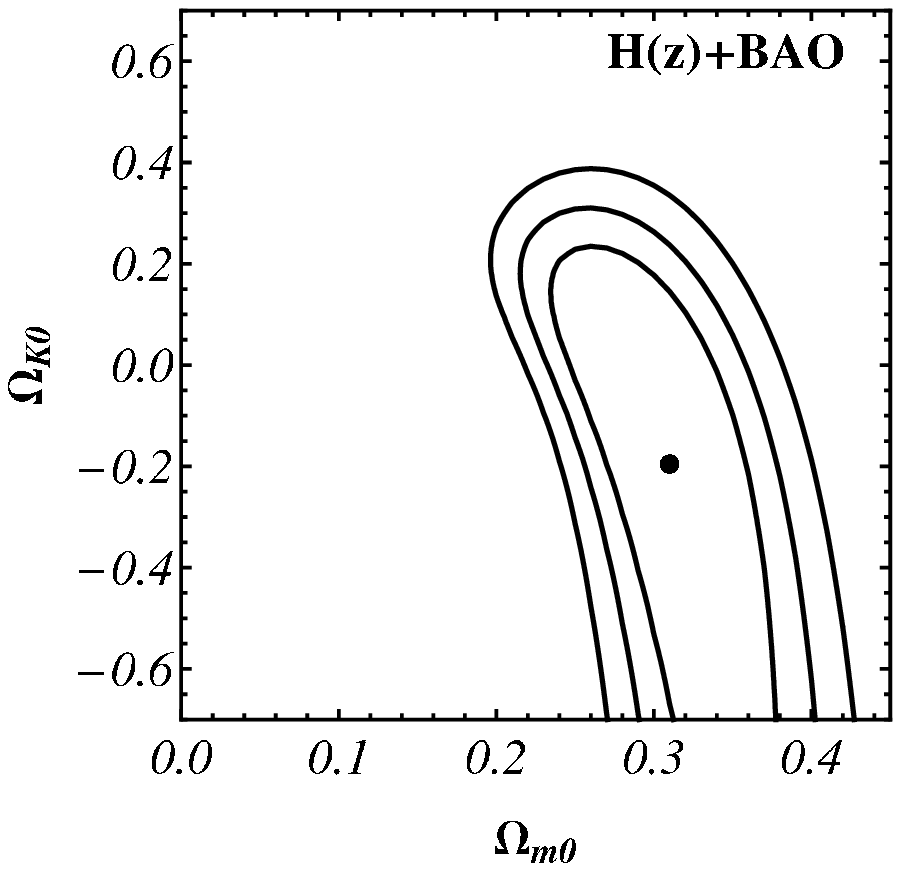}
  \includegraphics[width=50mm]{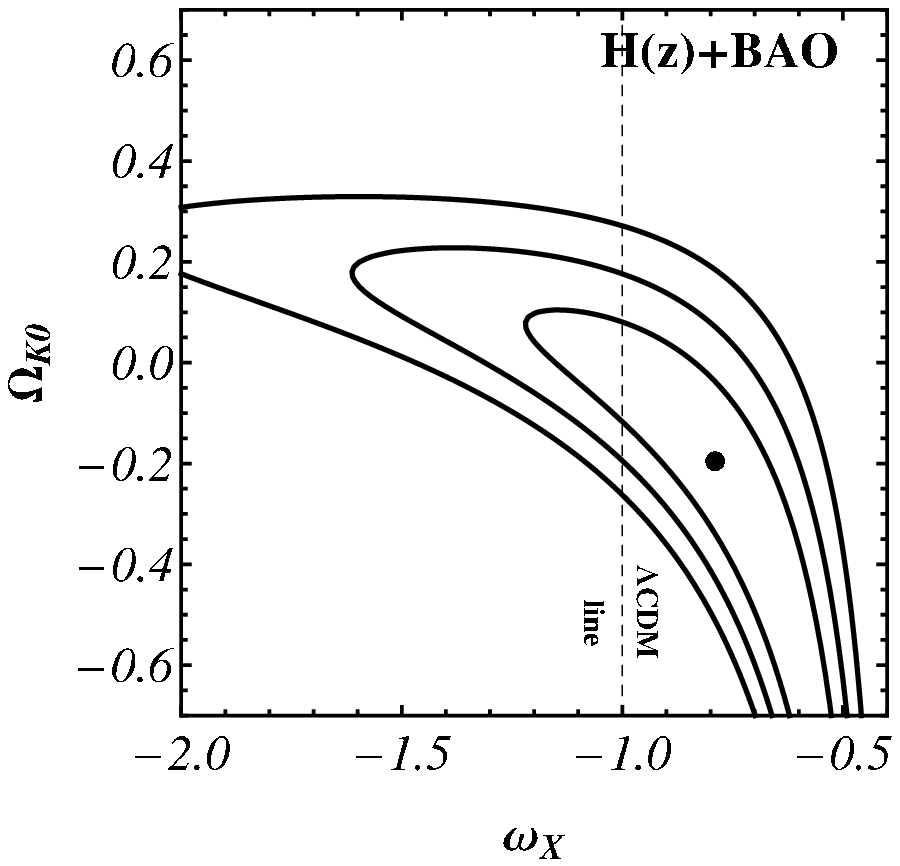}
  \includegraphics[width=50mm]{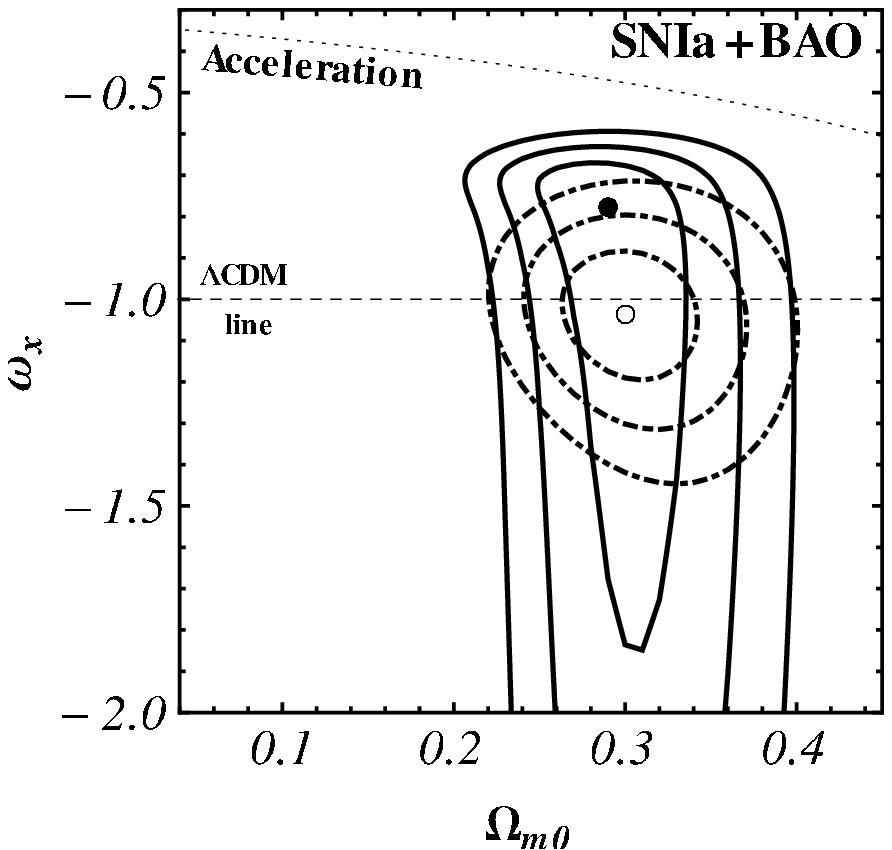}
  \includegraphics[width=50mm]{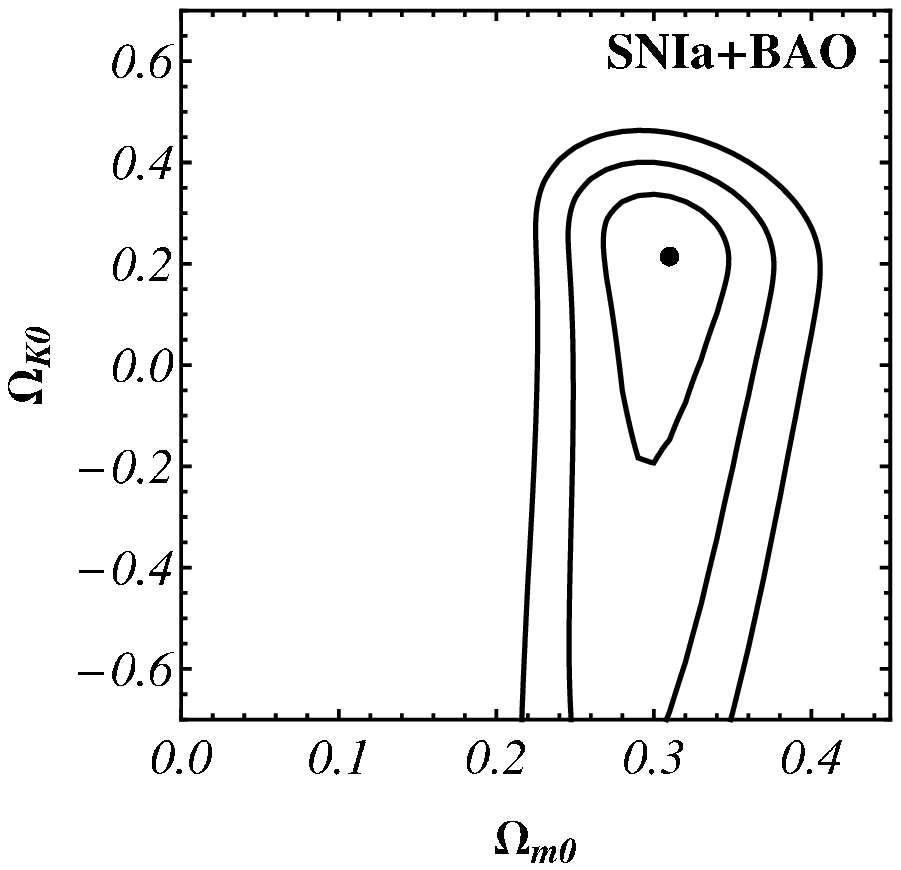}
  \includegraphics[width=50mm]{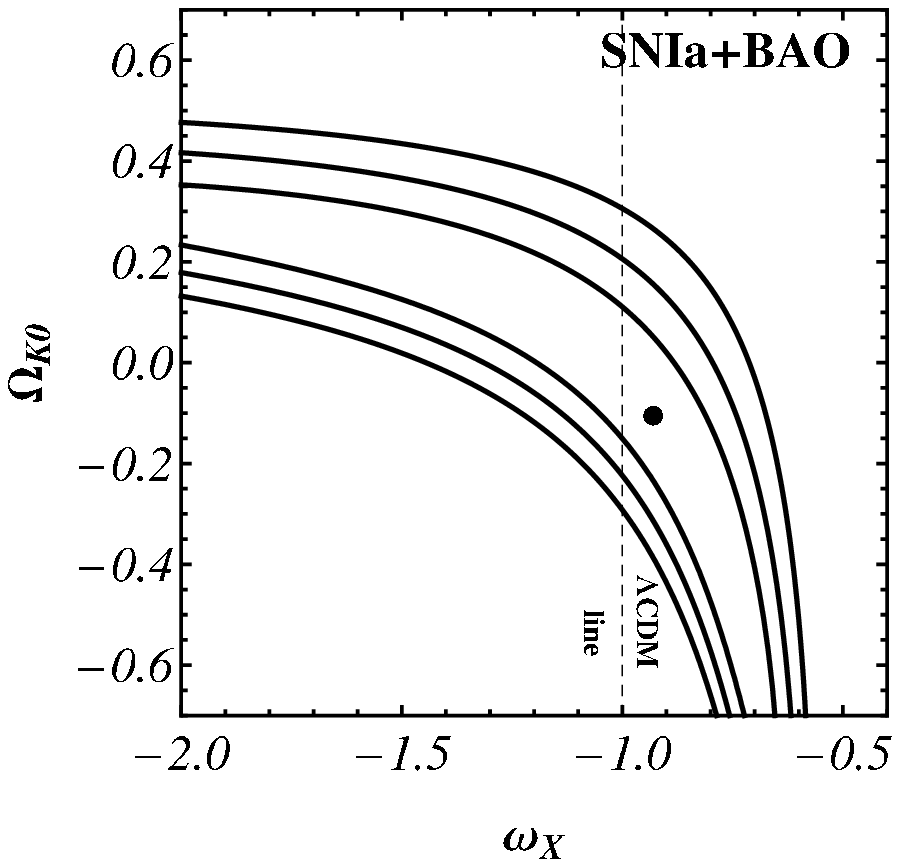}

\caption{
$1\sigma$, $2\sigma$, and $3\sigma$ constraint contours (solid lines) for parameters of the non-flat XCDM dark energy parameterization from 
$H(z)+$SNIa (first row), $H(z)+$BAO (second row), and SNIa$+$BAO (third row) measurements; filled circles show best-fit points. The dot-dashed lines in the first column panels are $1\sigma$, $2\sigma$, and $3\sigma$ constraint contours
derived by \cite{Farooq2013a} using the spatially-flat XCDM dark energy parameterization (open circles show best-fit points); here dotted lines distinguish between
accelerating and decelerating models (at zero space curvature) and dashed lines (here and in the third column) correspond to the $\Lambda$CDM model. First, second, and third columns correspond to marginalizing over $\Omega_{K0}$, 
$\omega_X$, and $\Omega_{m0}$ respectively.
} \label{fig:XCDM_D}
\end{figure}

\begin{figure}[p]
\centering
  \includegraphics[width=48mm]{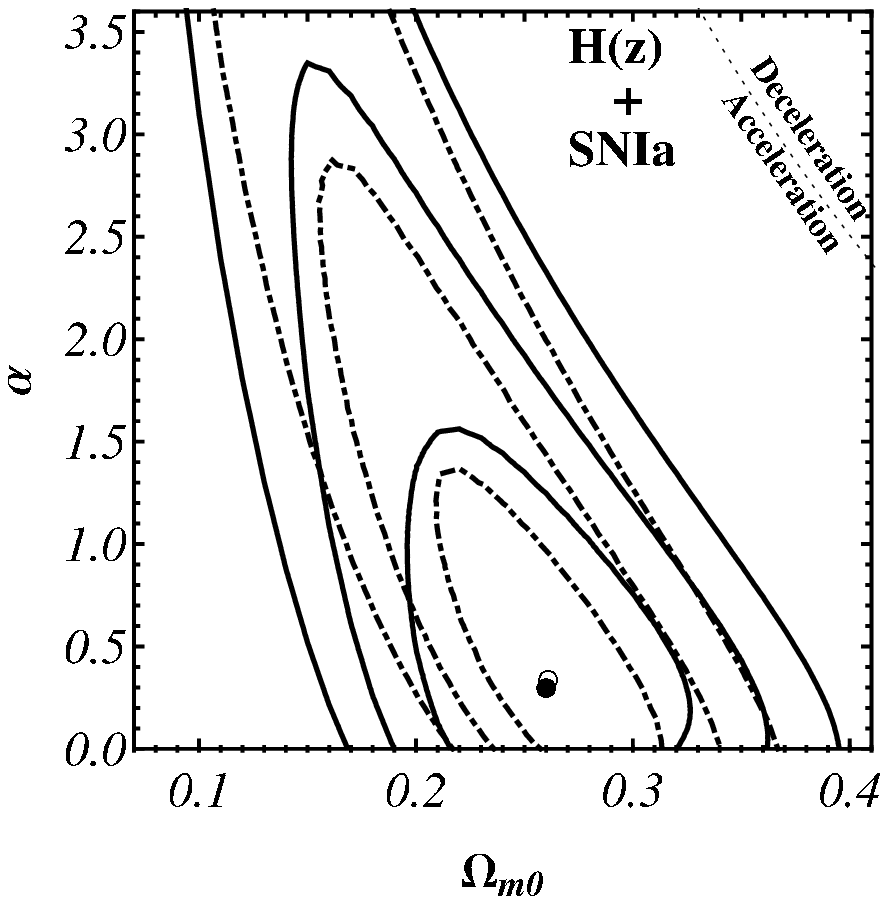}
  \includegraphics[width=50mm]{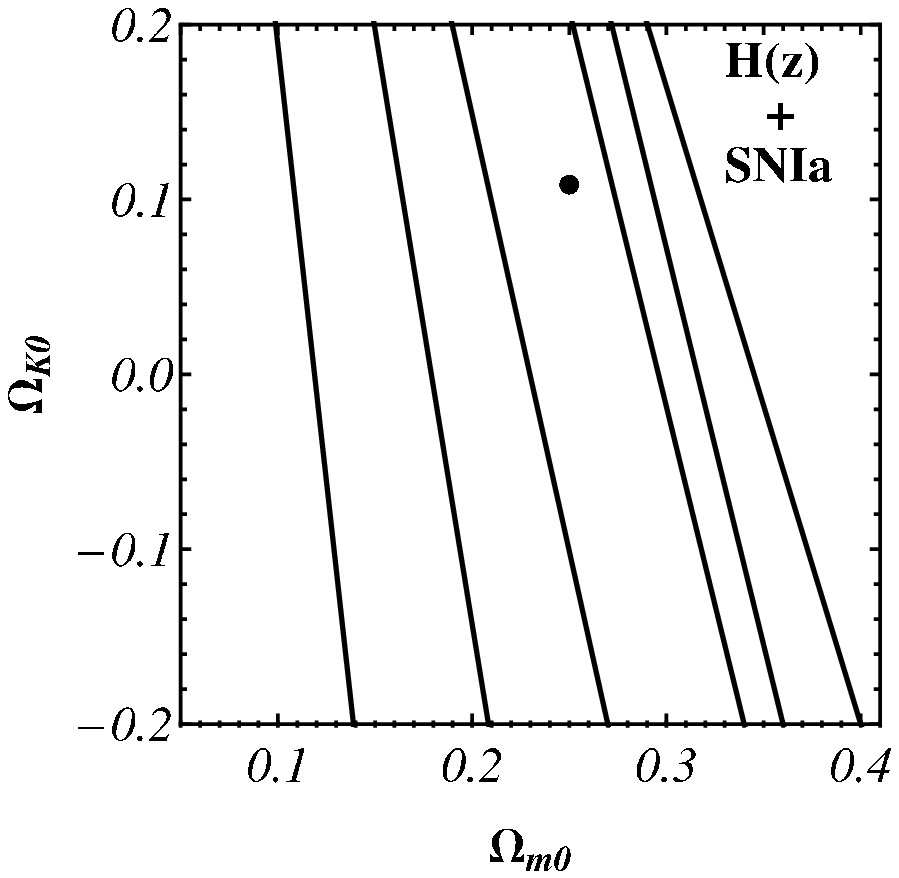}
  \includegraphics[width=51mm]{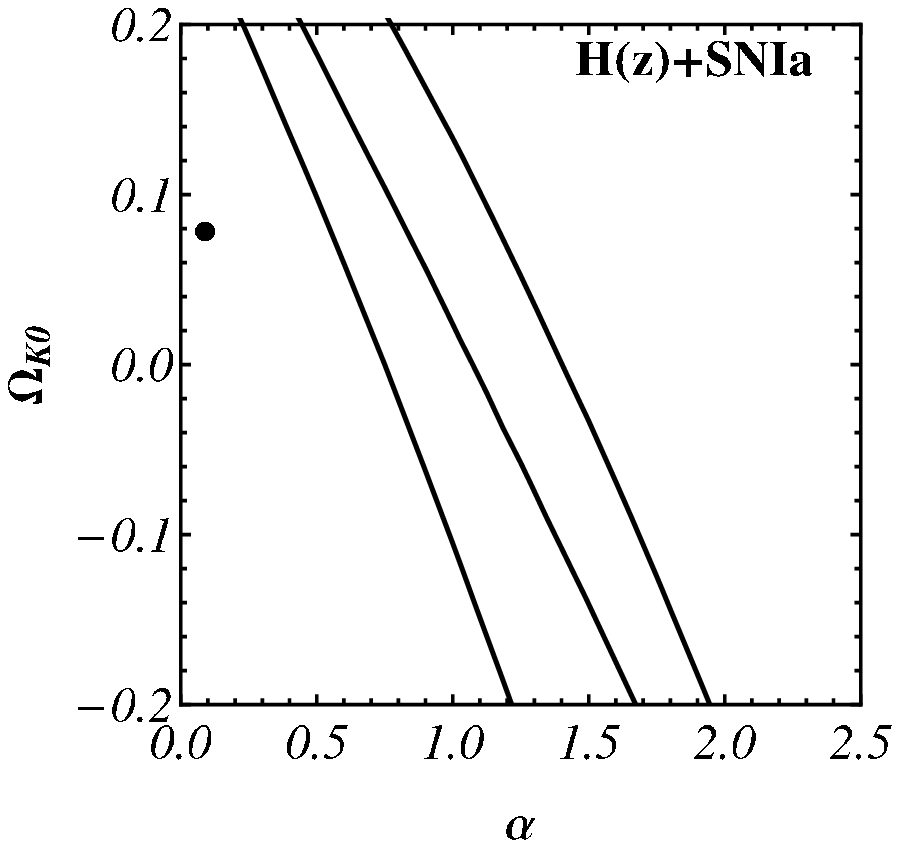}
  \includegraphics[width=48mm]{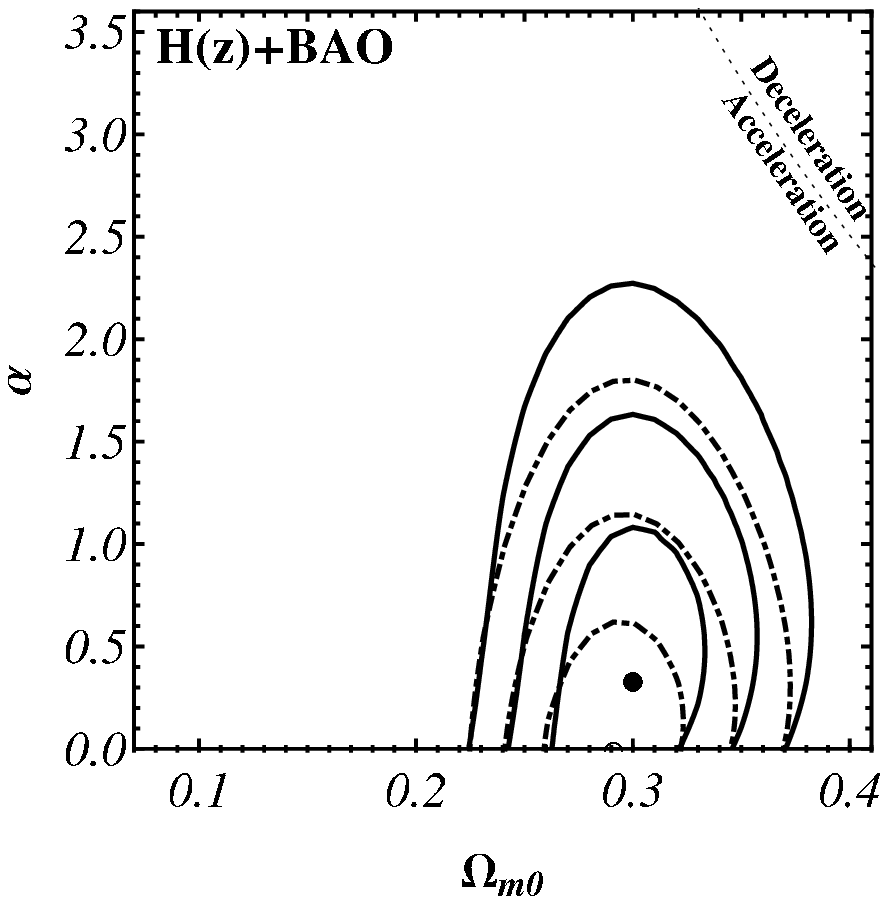}
  \includegraphics[width=50.5mm]{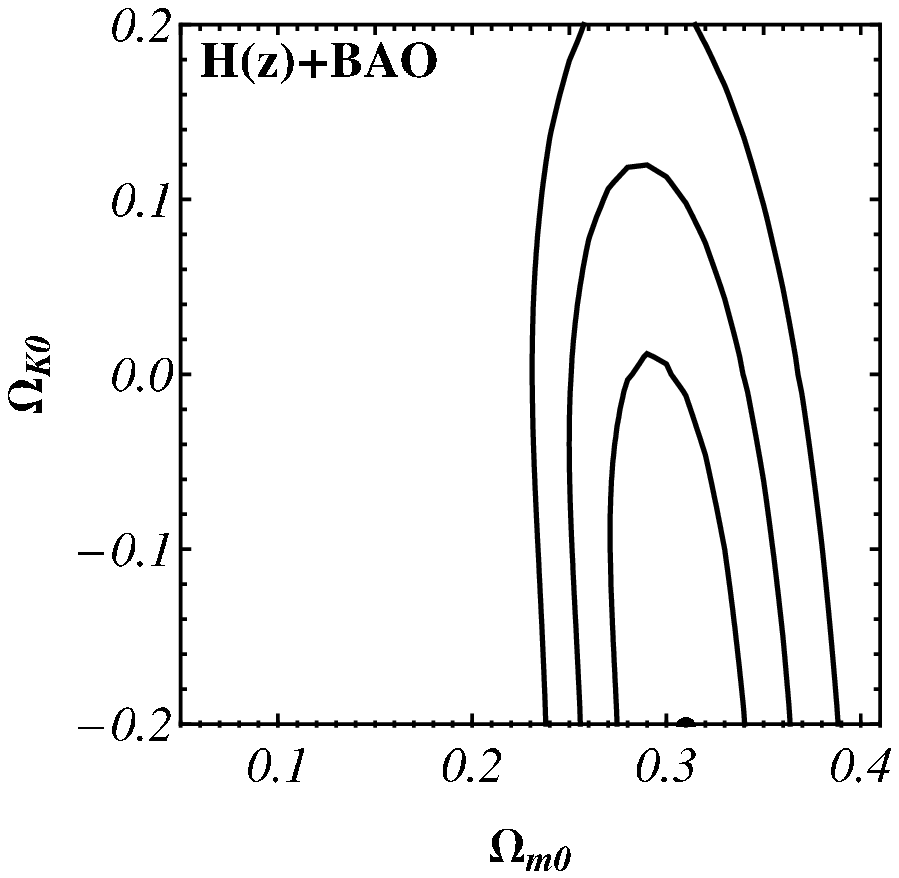}
  \includegraphics[width=51mm]{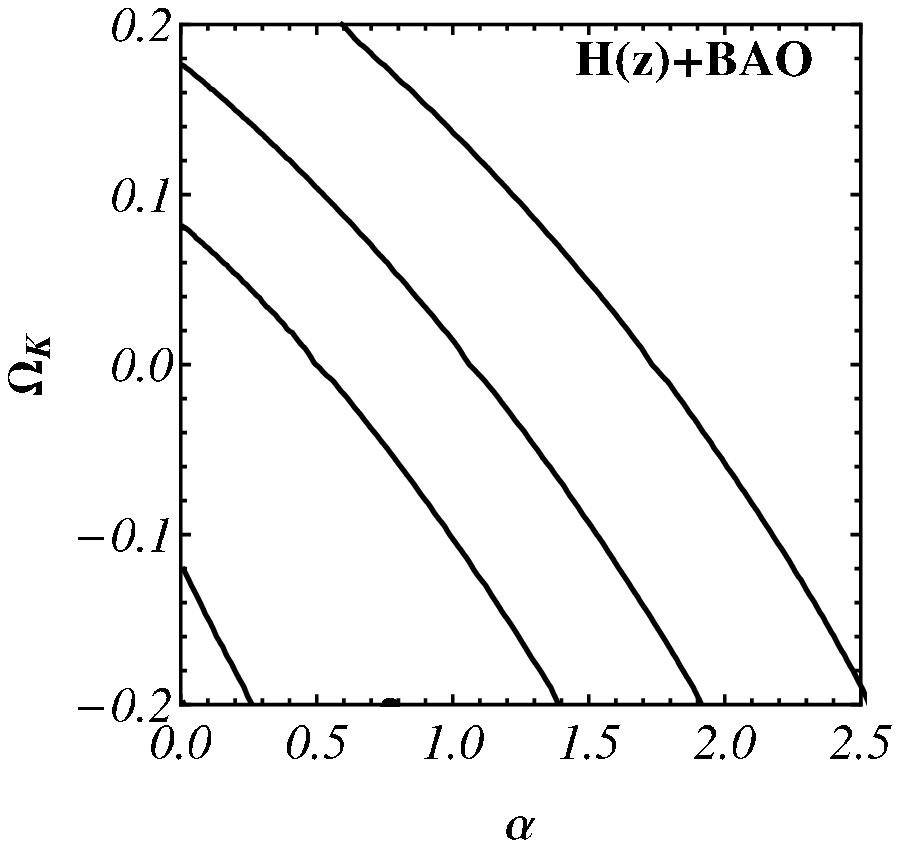}
  \includegraphics[width=48mm]{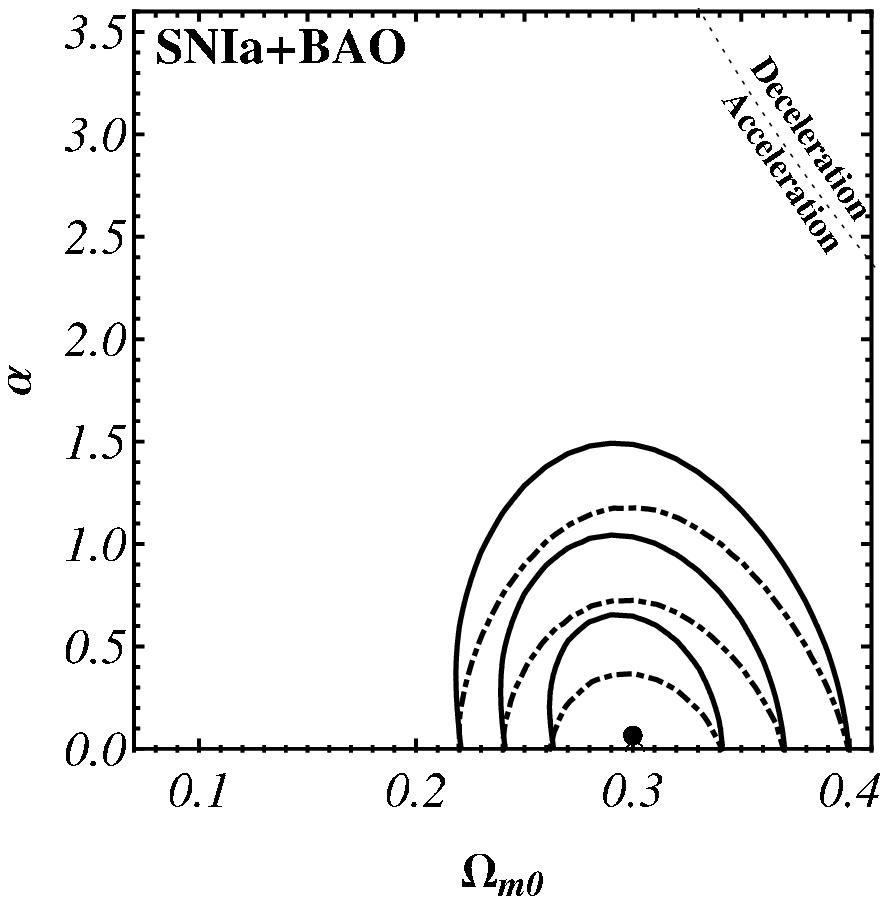}
  \includegraphics[width=50.5mm]{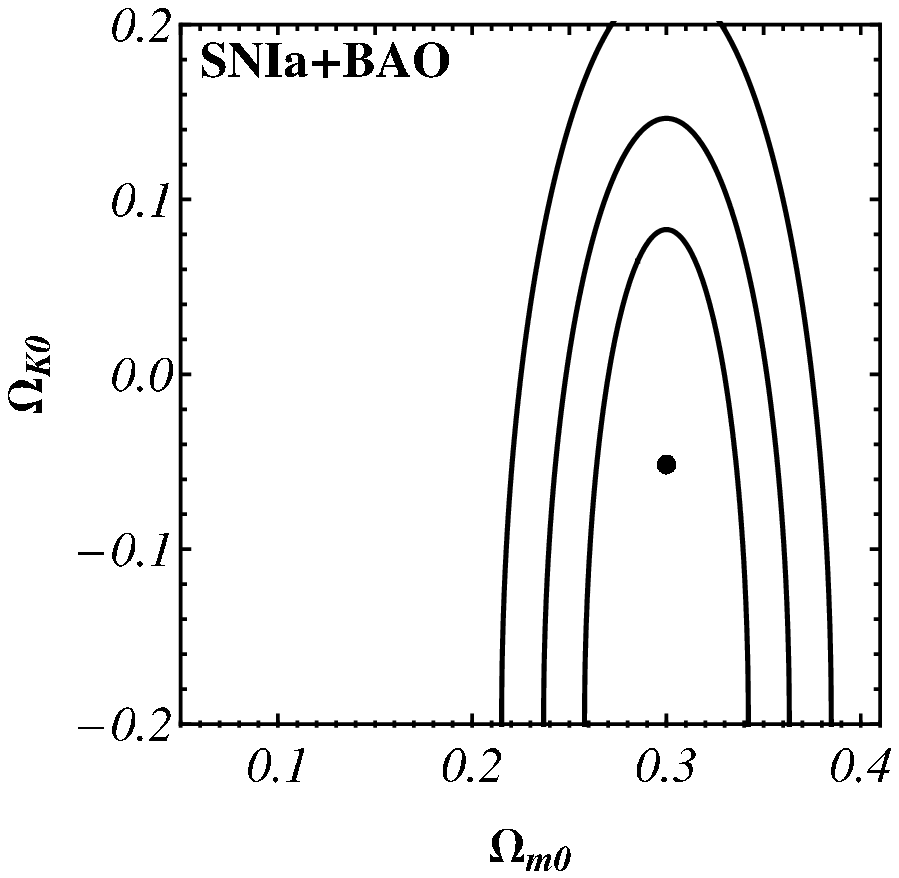}
  \includegraphics[width=51mm]{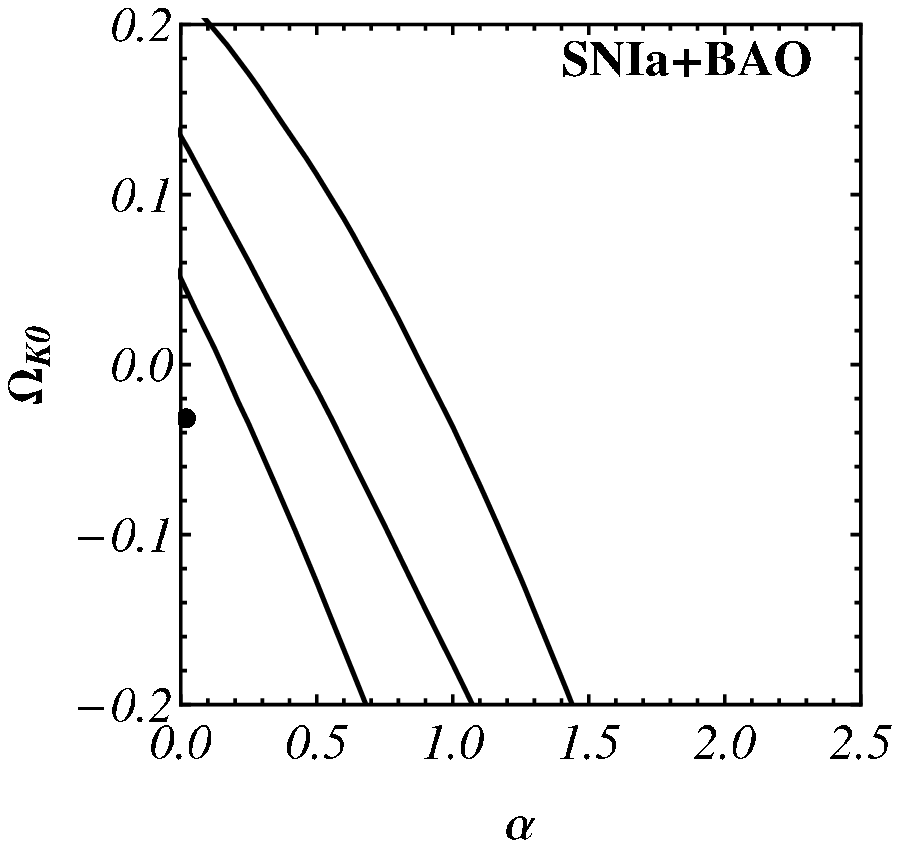}
\caption{
$1\sigma$, $2\sigma$, and $3\sigma$ constraint contour (solid lines) for parameters of the non-flat $\phi$CDM dark energy model from 
$H(z)+$SNIa (first row), $H(z)+$BAO (second row), and BAO$+$SNIa (third row) measurements; filled circles show best-fit points. The dot-dashed lines in the first column panels are $1\sigma$, $2\sigma$, and $3\sigma$ constraint contours
derived by \cite{Farooq2013a} using the spatially-flat $\phi$CDM model (open circles show best-fit points); here dotted lines distinguish between
accelerating and decelerating models (at zero space curvature) and the $\alpha=0$ axes (here and in the third column) correspond to the $\Lambda$CDM model. First, second, and third columns correspond to marginalizing over $\Omega_{K0}$, 
$\alpha$, and $\Omega_{m0}$ respectively. {We note that the $\Omega_{K0}=0$ constraints and the marginalized $\Omega_{K0}$ constraints differ by only a small amount due to the prior range of $\Omega_{K0}$ used here.} 
} \label{fig:phiCDM_D}
\end{figure}

Figures\ \ref{fig:XCDM_D} and
\ref{fig:phiCDM_D} show constraints on the parameters of the XCDM parameterization
and the $\phi$CDM model from the $H(z)$+SNIa (top row), $H(z)$+BAO (middle row), 
and SNIa+BAO (bottom row) measurements. In these figures the panels in the first, second, and
third columns show the two-dimensional probability density 
constraint contours (solid lines) from $\mathcal{L}(\Omega_{m0},\omega_X)
[\mathcal{L}(\Omega_{m0},\alpha)]$, $\mathcal{L}(\Omega_{m0},\Omega_{K0})$, and
$\mathcal{L}(\omega_X,\Omega_{K0})[\mathcal{L}(\alpha,\Omega_{K0})]$ for the
XCDM parameterization [the $\phi$CDM model]. The dot-dashed contours in the panels of
the first columns of Figs.\ \ref{fig:XCDM_D} and 
\ref{fig:phiCDM_D} are $1\sigma$, $2\sigma$, and $3\sigma$ confidence contours
corresponding to spatially-flat models, reproduced from \cite{Farooq2013a}.
Tables\ \ref{table:XCDMresults}
and \ref{table:phiCDMresults} list best-fit points and $\chi^2_{\mathrm{min}}$ values.

Comparing the solid contours of Figs.\ \ref{fig:XCDM_D} and \ref{fig:phiCDM_D} to those derived from the
single data sets in Figs.\ \ref{fig:XCDM_S} and \ref{fig:phiCDM_S}, we see that combinations of pairs of data
sets result in a significant tightening of constraints, especially on $\Omega_{m0}$, and less so on 
$\Omega_{K0}$, $\omega_{X}$, and $\alpha$. 

Comparing the solid contours to the dot-dashed contours in the panels in the first columns of 
Figs.\ \ref{fig:XCDM_D} and \ref{fig:phiCDM_D} we see that the addition of space curvature
as a third free parameter results in a fairly significant broadening of the 
constraint contours, even when using two data sets at a time, particularly in the direction along the parameter
that governs the time evolution of the dark energy density ($\omega_{X}$ for the XCDM parameterization and $\alpha$ for the $\phi$CDM
model). Again, when space curvature is included as a free parameter the constraint contours broaden more significantly for the
XCDM parameterization than for the $\phi$CDM model: compare the solid and dot-dashed contours in the first columns 
of Figs.\ \ref{fig:XCDM_D} and \ref{fig:phiCDM_D} (this is probably partially a consequence of the 
smaller range of space curvature, $-0.2 \leq \Omega_{K0} \leq 0.2$, considered for
computational tractability in the $\phi$CDM case).

\begin{figure}[p]
\centering
  \includegraphics[width=78.3mm]{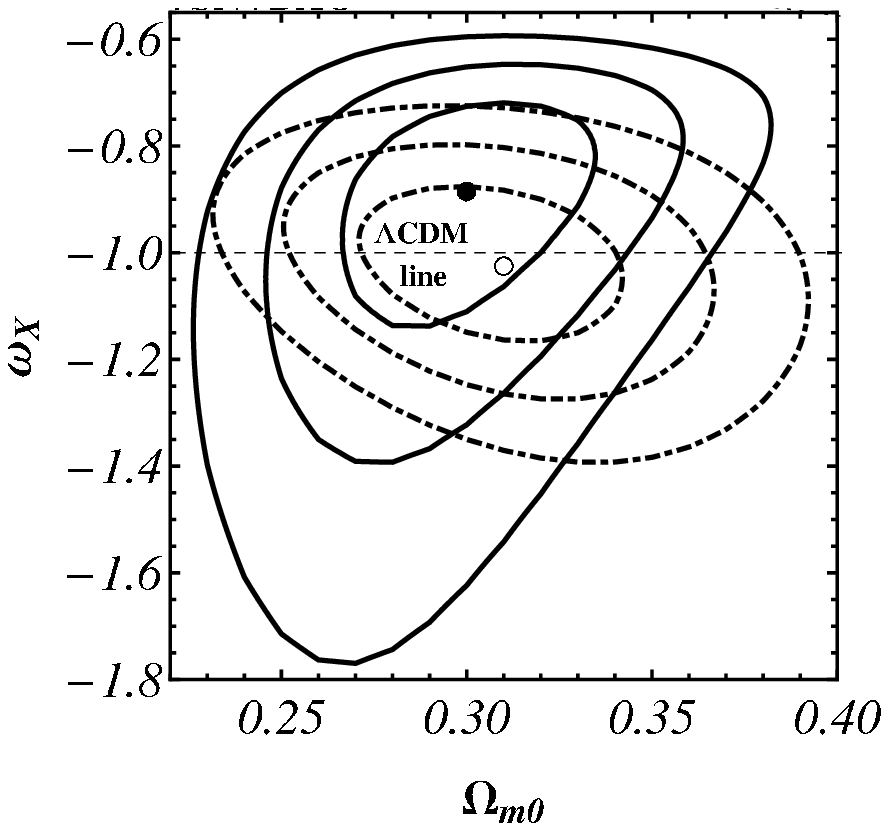}
  \includegraphics[width=74.0mm]{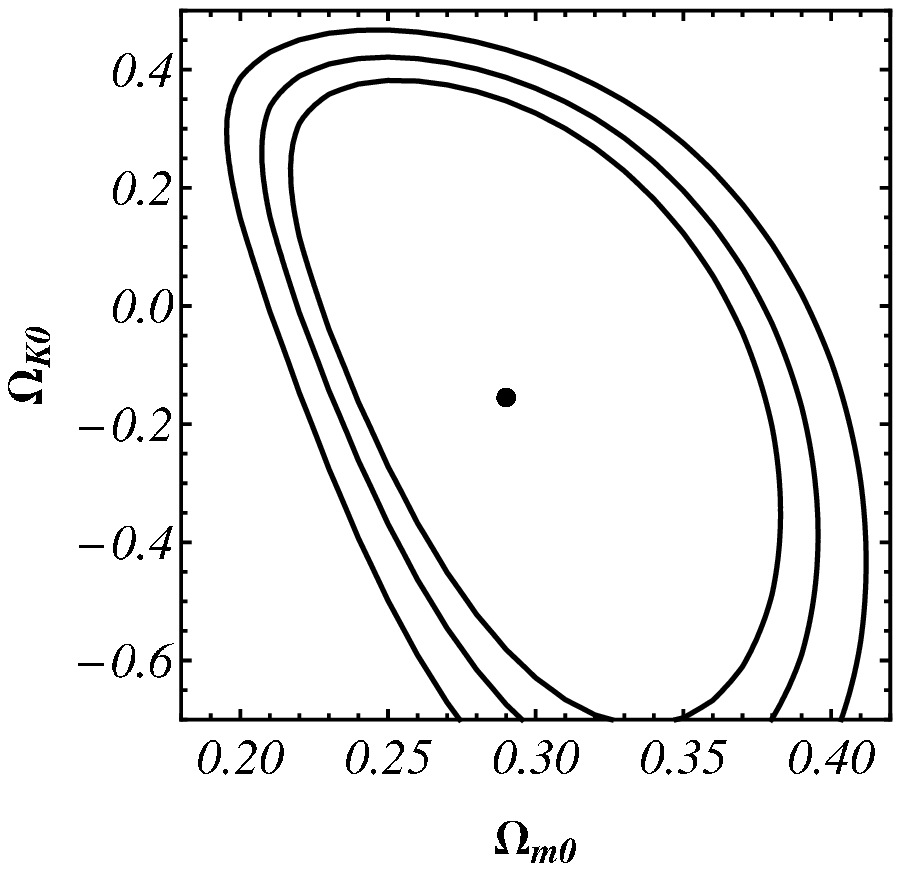}
  \includegraphics[width=74.0mm]{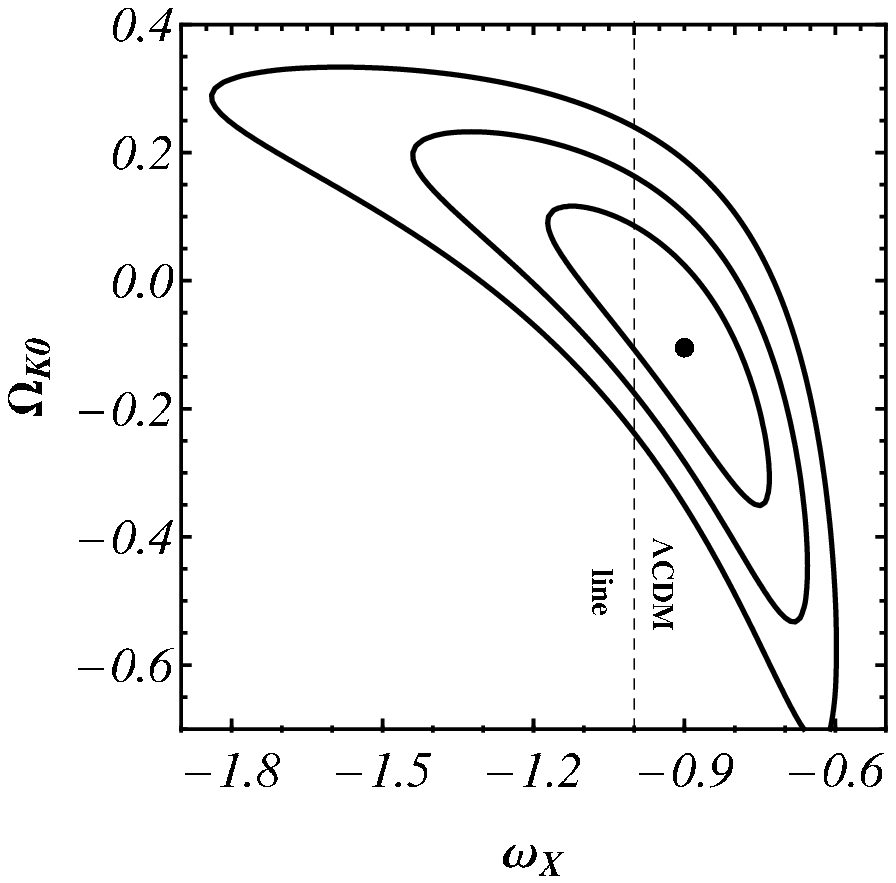}

\caption{
$1\sigma$, $2\sigma$, and $3\sigma$ constraint contours (solid lines) for parameters of the non-flat XCDM dark energy parameterization from 
$H(z)+$SNIa$+$BAO measurements; filled circles show best-fit points. The dot-dashed lines in the top left panel are $1\sigma$, $2\sigma$, and $3\sigma$ constraint contours
derived by \cite{Farooq2013a} using the spatially-flat XCDM parameterization (open circle shows best-fit point); here dashed lines (in the top left and bottom panels) correspond to the $\Lambda$CDM model. Top left, top right, and bottom panel correspond to marginalizing over $\Omega_{K0}$, 
$\omega_X$, and $\Omega_{m0}$ respectively.
} \label{fig:XCDM_A}
\end{figure}

\begin{figure}[p]
\centering
  \includegraphics[width=75.0mm]{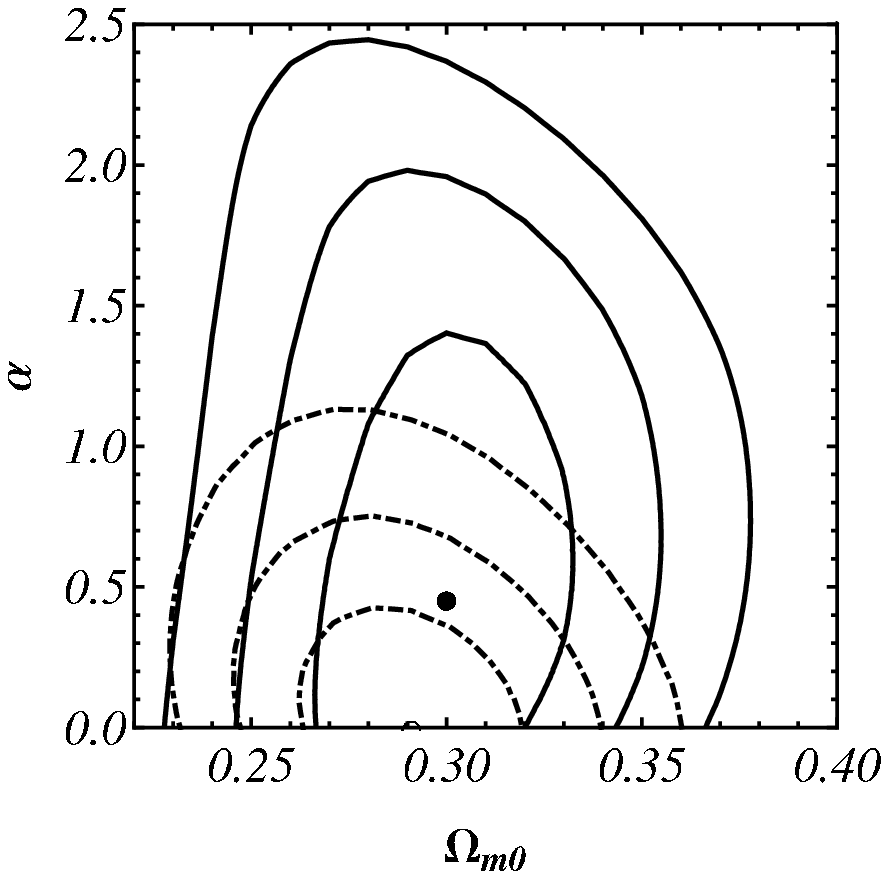}
  \includegraphics[width=75.0mm]{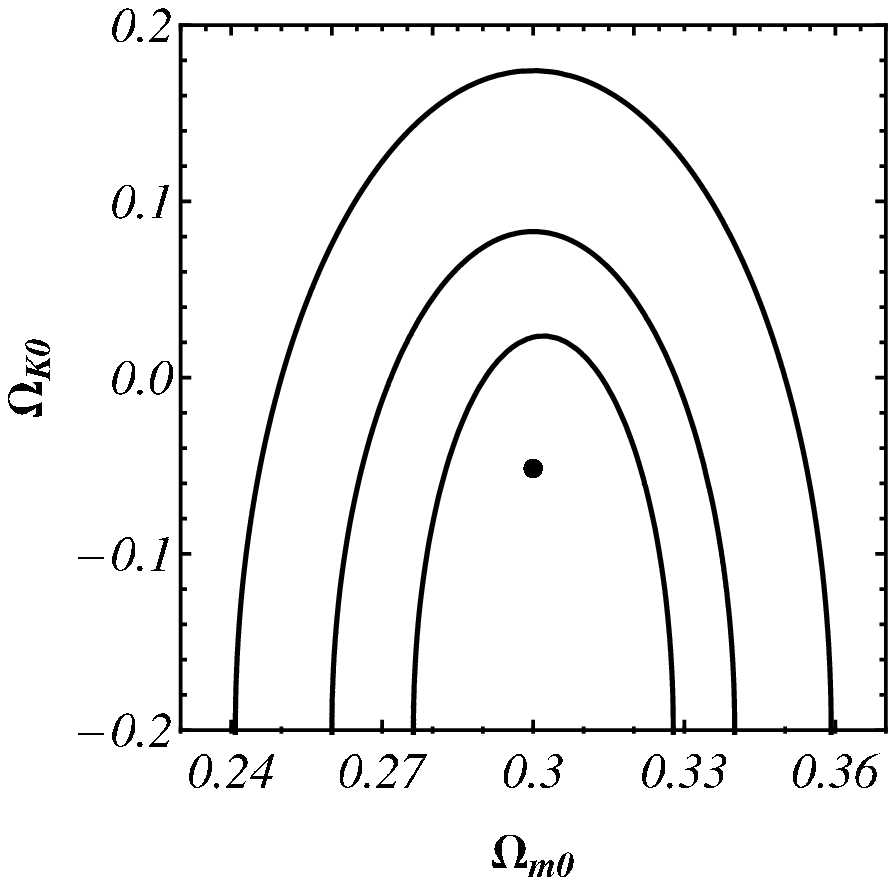}
  \includegraphics[width=75.0mm]{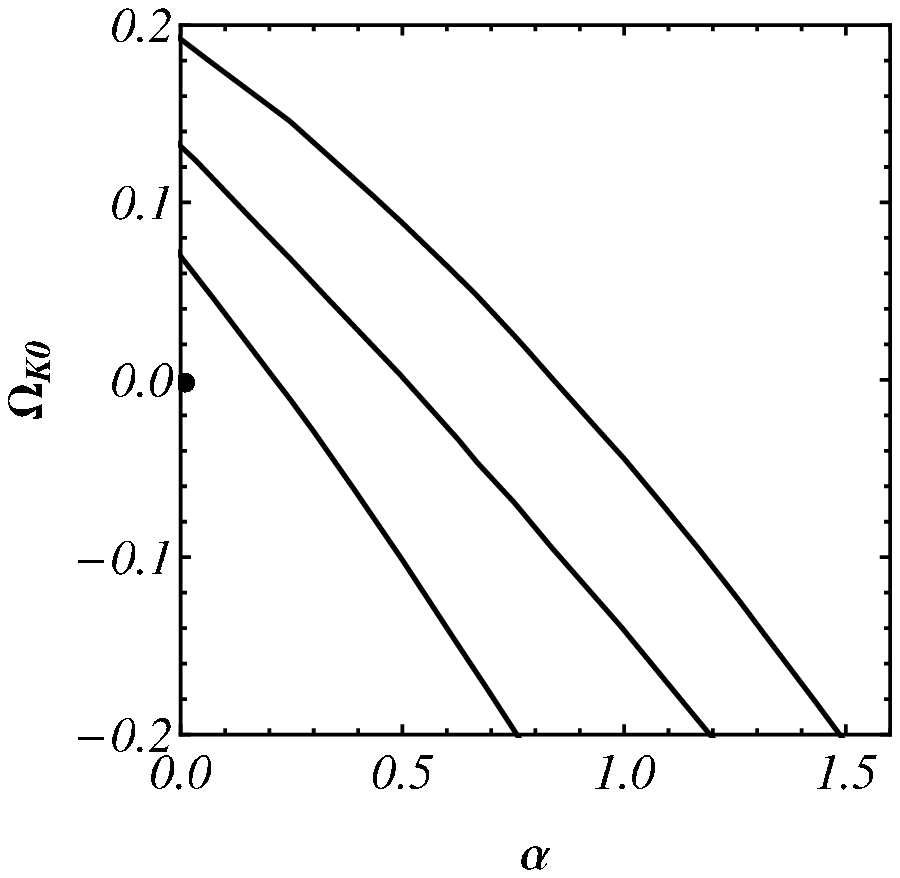}
\caption{
$1\sigma$, $2\sigma$, and $3\sigma$ constraint contours (solid lines) for parameters of the non-flat $\phi$CDM dark energy model from
$H(z)+$SNIa$+$BAO measurements; filled circles show best-fit points. The dot-dashed lines are $1\sigma$, $2\sigma$, and $3\sigma$ constraint contours
derived by \cite{Farooq2013a} using the spatially-flat $\phi$CDM model (open circle shows best-fit point); here the  $\alpha=0$ axes in the top left and bottom panels correspond to the $\Lambda$CDM model. Top left, top right, and bottom panel correspond to marginalizing over $\Omega_{K0}$, 
$\alpha$, and $\Omega_{m0}$ respectively.
} \label{fig:phiCDM_A}
\end{figure}

Encouraged by the tightening of the constraint contours when two data sets are analyzed together, we now discuss the result of a joint analysis of the $H(z)$, SNIa, and BAO data. 
Figures\ \ref{fig:XCDM_A} and
\ref{fig:phiCDM_A} show constraints on the parameters of the XCDM parameterization
and the $\phi$CDM model from the $H(z)$+SNIa+BAO measurements. 
In these figures the top left panel, top right panel, and the bottom panel
show the two-dimensional probability density 
constraint contours (solid lines) from $\mathcal{L}(\Omega_{m0},\omega_X)
[\mathcal{L}(\Omega_{m0},\alpha)]$, $\mathcal{L}(\Omega_{m0},\Omega_{K0})$, and
$\mathcal{L}(\omega_X,\Omega_{K0})[\mathcal{L}(\alpha,\Omega_{K0})]$ for the
XCDM parameterization [the $\phi$CDM model]. The dot-dashed contours in the left top panels of
Figs.\ (\ref{fig:XCDM_A}) and 
(\ref{fig:phiCDM_A}) are $1\sigma$, $2\sigma$, and $3\sigma$ confidence contours
corresponding to spatially-flat models, reproduced from \cite{Farooq2013a}.
Tables\ \ref{table:XCDMresults}
and \ref{table:phiCDMresults} list best-fit points and $\chi^2_{\mathrm{min}}$ values.

Comparing the solid contours of Figs.\ \ref{fig:XCDM_A} and \ref{fig:phiCDM_A} to those derived from the
data set pairs of Figs.\ \ref{fig:XCDM_D} and \ref{fig:phiCDM_D}, we see that the joint analysis of all three data sets results in a significant tightening of constraints.

Comparing the solid contours to the dot-dashed contours in the left top panels in 
Figs.\ \ref{fig:XCDM_A} and \ref{fig:phiCDM_A} we see that the addition of space curvature
as a third free parameter results in a significant broadening of the 
constraint contours, but this time less than when only two data sets were used in Figs.\ \ref{fig:XCDM_D} 
and \ref{fig:phiCDM_D}. The broadening is more significant in the direction along the parameter
that governs the time evolution of the dark energy density ($\omega_{X}$ for the XCDM parameterization and $\alpha$ for the $\phi$CDM
model).

We also computed the $1\sigma$ and $2\sigma$ bounds on model parameters that follow from the joint analysis 
of $H(z)$, SNIa, and BAO measurements. Tables\ \ref{table:XCDMintervals} and \ref{table:phiCDMintervals} 
list these bounds on
individual cosmological parameters, determined from their one-dimensional posterior probability distribution 
functions (which we obtained by marginalizing the three-dimensional likelihood over the other two cosmological parameters). The numerical values listed in these tables confirm the results described in the discussion above of Figs.\ \ref{fig:XCDM_A} and \ref{fig:phiCDM_A}.


\begin{center}
\begin{threeparttable}
\caption{XCDM Parametrization Results From $H(z)+$SNIa$+$BAO Data}
\begin{tabular}{ccc}

\hline\hline

\multirow{1}{*}{Marginalization Range} &  1$\sigma$ intervals & 2$\sigma$ intervals\\
\hline
\multirow{2}{*}{$\Omega_{K0} =0$\tnote{a}} & $0.29\leqslant \Omega_{m0} \leqslant 0.31$ & $0.27\leqslant \Omega_{m0} \leqslant 0.32$\\
{}& $-1.01\leqslant \omega_{X} \leqslant -0.83$ & $-1.03 \leqslant \omega_{X} \leqslant -0.77$\\
\cline{1-3}
\multirow{2}{*}{$-0.7 \leq \Omega_{K0}\leq -0.7$} & $0.27\leqslant \Omega_{m0} \leqslant 0.32$ & $0.25\leqslant \Omega_{m0} \leqslant 0.34$\\
{}& $-1.03\leqslant \omega_{X} \leqslant -0.77$ & $-1.25 \leqslant \omega_{X} \leqslant -0.69$\\
\cline{1-3}
\multirow{2}{*}{$-2 \leq \omega_{X}\leq 0$} & $0.27\leqslant \Omega_{m0} \leqslant 0.32$ & $0.25\leqslant \Omega_{m0} \leqslant 0.34$\\
{}& $-0.21\leqslant \Omega_{K0} \leqslant 0.10$ & $-0.39 \leqslant \Omega_{K0} \leqslant 0.22$\\
\cline{1-3}
\multirow{2}{*}{$0 \leq \Omega_{m0}\leq 1$} & $-1.03\leqslant \omega_{X} \leqslant -0.77$ & $-1.30\leqslant \omega_{X} \leqslant -0.69$\\
{}& $-0.21\leqslant \Omega_{K0} \leqslant 0.10$ & $-0.39 \leqslant \Omega_{K0} \leqslant 0.22$\\
\cline{1-3}
\end{tabular}
\begin{tablenotes}
\item[a]{From \cite{Farooq2013a}.}
\end{tablenotes}
\label{table:XCDMintervals}
\end{threeparttable}
\end{center}


\begin{center}
\begin{threeparttable}
\caption{$\phi$CDM Model Results From $H(z)+$SNIa$+$BAO Data}

\begin{tabular}{ccc}

\hline\hline

\multirow{1}{*}{Marginalization Range} &  1$\sigma$ intervals & 2$\sigma$ intervals\\
\hline
\multirow{2}{*}{$\Omega_{K0} =0$\tnote{a}} & $0.27\leqslant \Omega_{m0} \leqslant 0.29$ & $0.25\leqslant \Omega_{m0} \leqslant 0.30$\\
{}& $\alpha \leqslant 0.31$ & $ \alpha \leqslant 0.56$\\
\cline{1-3}
\multirow{2}{*}{$-0.2 \leq \Omega_{K0}\leq -0.2$} & $0.28\leqslant \Omega_{m0} \leqslant 0.32$ & $0.26\leqslant \Omega_{m0} \leqslant 0.34$\\
{}& $\alpha  \leqslant 1.03$ & $ \alpha \leqslant 1.64$\\
\cline{1-3}
\multirow{2}{*}{$0 \leq \alpha\leq 5$\tnote{b}} & $0.28\leqslant \Omega_{m0} \leqslant 0.31$ & $0.26\leqslant \Omega_{m0} \leqslant 0.33$\\
{}& $-0.2\leqslant \Omega_{K0} \leqslant 0.09$ & $-0.2 \leqslant \Omega_{K0} \leqslant 0.12$\\
\cline{1-3}
\multirow{2}{*}{$0 \leq \Omega_{m0}\leq 1$\tnote{b}} & $\alpha  \leqslant 1.03$ & $ \alpha \leqslant 1.64$\\
{}& $-0.2\leqslant \Omega_{K0} \leqslant 0.09$ & $-0.2 \leqslant \Omega_{K0} \leqslant 0.12$\\
\cline{1-3}
\end{tabular}
\begin{tablenotes}
\item[a]{From \cite{Farooq2013a}.}
\item[b]{The lower limit on $\Omega_{K0}$ is determined by the lower limit of the $\Omega_{K0}\geq -0.2$ prior assumed for the case of $\phi$CDM, not from the observational data. However, we strongly suspect that this limit will not change greatly with an increase of the integration limit for the data used here.}
\end{tablenotes}
\label{table:phiCDMintervals}
\end{threeparttable}
\end{center}

Of some interest are the bounds on the curvature density parameter
$\Omega_{K0}$. Perhaps the most useful summary is the 1$\sigma$ limit 
$|\Omega_{K0}|\lesssim  0.15$ derived by symmetrizing about $\Omega_{K0}=0$
the 1$\sigma$ range from the central columns of Table \ref{table:XCDMintervals} and
\ref{table:phiCDMintervals}. Note that the possible 2$\sigma$ range of $\Omega_{K0}$ is
significantly smaller for $\phi$CDM than for XCDM (compare the relevant entries in 
the last columns of Table \ref{table:XCDMintervals} and
\ref{table:phiCDMintervals}). This is almost certainly a consequence of the
smaller range of $\Omega_{K0}$, $-0.2 \leq \Omega_{K0} \leq 0.2$, we have used in
the $\phi$CDM computation; the 2$\sigma$ XCDM bound $|\Omega_{K0}|\lesssim  0.3$ is the
more reliable one.

\section{Conclusion}
\label{summary}

A joint analysis of $H(z)$, SNIa, and BAO data using the XCDM parametrization and the $\phi$CDM model
of time evolving dark energy density
in a non-flat geometry leads to the conclusion that more, and more precise, data are required to tightly
pin down the spatial curvature of the Universe in dynamical dark energy models. These data require 
$|\Omega_{K0}|\lesssim  0.15$ at 1$\sigma$ confidence. It would be of interest
to determine the constraints on space curvature in the non-flat $\phi$CDM model
from CMB anisotropy measurements. Such an analysis, possibly in combination with 
that of other data of the kind considered here, and extended over a wider range of
$\Omega_{K0}$ than we have considered, could go a long way towards establishing whether
space curvature contributes significantly to the current cosmological energy budget.

\acknowledgments

We thank Mikhail Makouski, Anatoly Pavlov, and Shawn Westmoreland for 
useful discussions and helpful advice. We thank Daniel Nelson
for allowing us to use his computer for some of the computations.
This work was supported in part by DOE grant DEFG03-99EP41093 
and NSF grant AST-1109275.


\def\mnras{MNRAS}
\def\aapr{A\&A~Rev.}
\def\jcap{J. Cosmology Astropart. Phys.}
\def\apjl{ApJ}
\def\pasp{PASP}
\def\aap{A\&A}
\def\apss{Ap\&SS}
\def\apjs{ApJS}
\def\apj{ApJ}
\def\prd{Phys. Rev. D}

\end{document}